\documentclass[%
  aps,%
  pra,%
  groupedaddress,%
  10pt,%
  twocolumn,%
]{revtex4-2}

\usepackage[utf8]{inputenc}
\usepackage[english]{babel}
\usepackage[caption=false]{subfig}

\usepackage{graphicx}
\usepackage{siunitx}
\usepackage{amsmath, amssymb}
\usepackage{mathrsfs}
\usepackage{upgreek}
\usepackage{chemmacros}
\usepackage{xspace}
\usepackage{bm}
\usepackage{calc}
\usepackage{comment}
\usepackage{moreverb}
\usepackage{setspace}
\usepackage{flafter}
\usepackage{float}
\usepackage{todonotes}
\usepackage{tikz}

\usetikzlibrary{
  arrows,
  decorations.pathreplacing,
  calligraphy}

\usepackage{etoolbox}
\robustify{\subref}

\graphicspath{{./}{./img/}{../img/}}

\newcommand*\diff{\mathop{}\!\mathrm{d}}

\DeclareMathOperator{\tr}{tr}

\usepackage{bbold}
\newcommand\ident{{\ensuremath\mathbb{1}}}

\newcommand\mawr[1]{\ensuremath{#1}\xspace}

\newcommand\ket[1]{\mawr{|#1\rangle}}

\newcommand\figref[1]{Fig.~\ref{fig:#1}}
\newcommand\subfigref[1]{Fig.~\ref{subfig:#1}}
\newcommand\eqnref[1]{Eq.~\ref{eqn:#1}}

\newcommand\tauLR{\tau_{\text{L}\rightarrow\text{R}}}
\newcommand\tauRL{\tau_{\text{R}\rightarrow\text{L}}}

\newcommand\tautr{\tau_{\text{tr}}}
\newcommand\taui{\tau_i}

\newcommand\PLL{P_{\text{L}\rightarrow\text{L}}}
\newcommand\PLR{P_{\text{L}\rightarrow\text{R}}}
\newcommand\PRL{P_{\text{R}\rightarrow\text{L}}}
\newcommand\PRR{P_{\text{R}\rightarrow\text{R}}}
\newcommand\Pmn{P_{\text{m}\rightarrow\text{n}}}
\newcommand\Ptr{P_{\text{tr}}}

\newcommand\invtczeno{\SI{3e17}{\per\s}\xspace}
\newcommand\invtcdiff{\SI{3e15}{\per\s}\xspace}
\newcommand\invtcnon{\SI{5e12}{\per\s}\xspace}

\newcommand\lratValidTrajectories{\num{6.4e6}\xspace}
\newcommand\lratDiscardedTrajectories{\num{448}\xspace}
\newcommand\sdotValidTrajectories{\num{2.9e5}\xspace}

\newcommand\psiloci{\psi_{\text{loc}}^i}
\newcommand\psilocperp{\psi_{\text{loc}}^\perp}

\newcommand\cD{\mathcal{D}}
\newcommand\sx{\mawr{\sigma_x}}

\definecolor{tabblue}{HTML}{1f77b4}
\definecolor{taborange}{HTML}{ff7f0e}
\definecolor{tabred}{HTML}{d62728}
\definecolor{tabgreen}{HTML}{2ca02c}
\definecolor{tabcyan}{HTML}{17becf}
\definecolor{tabolive}{HTML}{bcbd22}
\definecolor{tabpurple}{HTML}{9467bd}

\begin{document}

\title{Decoherence Effects Break Reciprocity in Matter Transport}

\author{P. Bredol}
\author{H. Boschker}
\author{D. Braak}
\author{J. Mannhart}
\email{corresponding author, office-mannhart@fkf.mpg.de}
\affiliation{Max Planck Institute for Solid State Research,
             70569 Stuttgart, Germany}

\date{\today}

\begin{abstract}
  The decoherence of quantum states defines the transition between the quantum world and classical physics. Decoherence or, analogously, quantum mechanical collapse events pose fundamental questions regarding the interpretation of quantum mechanics and are technologically relevant because they limit the coherent information processing performed by quantum computers. We have discovered that the transition regime enables a novel type of matter transport. Applying this discovery, we present nanoscale devices in which decoherence, modeled by random quantum jumps, produces fundamentally novel phenomena by interrupting the unitary dynamics of electron wave packets. Noncentrosymmetric conductors with mesoscopic length scales act as two-terminal rectifiers with unique properties.
  In these devices, the inelastic interaction of itinerant electrons with impurities acting as electron trapping centers leads to a novel steady state characterized by partial charge separation between the two leads, or, in closed circuits to the generation of persistent currents.
  The interface between the quantum and the classical worlds therefore provides a novel transport regime of value for the realization of a new category of mesoscopic electronic devices.
\end{abstract}

\maketitle
\thispagestyle{empty}

\section{Introduction}

The measurement process is a mysterious phenomenon at the heart of quantum physics. Starting with the Copenhagen interpretation \cite{can-quant-descr-of,heisenberg_physik_2000}, numerous approaches have attempted to describe this phenomenon. For an overview, see, e.g. \cite{london1983quantum}. The Copenhagen interpretation, for example, states that a measurement causes a spontaneous collapse of the wave function \cite{von2013mathematische}, whereas the decoherence theory attributes the apparent collapse to the entanglement of the system with its environment \cite{on-the-inter-of,decoh-einse-and-the,relat-state-formu-of}. Hitherto unknown processes are also considered to cause physically real quantum collapses, which for macroscopic systems create an observer-independent reality \cite{unifi-dynam-for}. In this work, we assume that physically real decoherence or collapse events are initiated by inelastic interactions and show that they impact mesoscopic transport without a macroscopic measurement process. This concept underlies the quantum trajectories approach widely and successfully used in quantum optics
\cite{quant-measu-and,quant-stoch-proce-as,wavef-appro-to-dissi,carmichael2009open,quant-traje-and-open,molmer1993}. We shall show that this method, applied to solid state systems on the nanoscale, predicts nonreciprocal transport properties of electrons.

\begin{figure}
\centering
\includegraphics[width=\linewidth,trim={0cm 10cm 0em 5cm},clip]{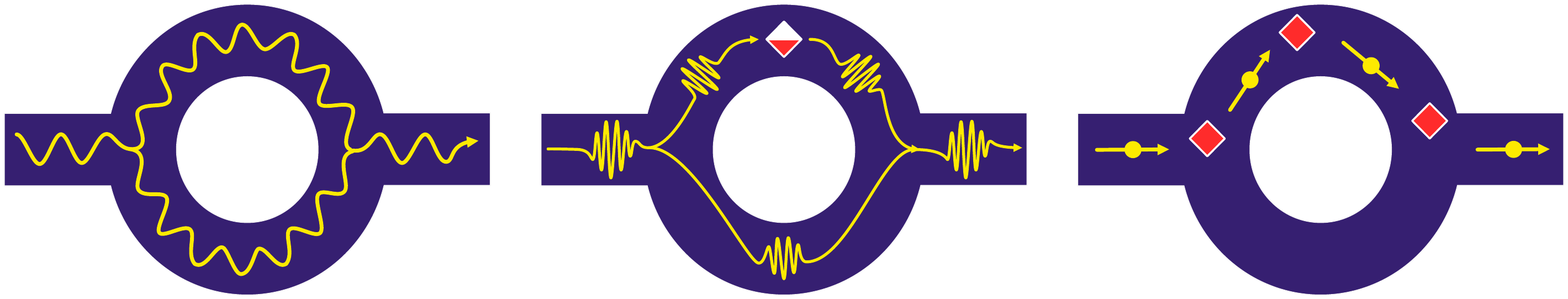}\\[-1em]
\subfloat[\label{subfig:quant}]{\hspace{.32\linewidth}}
\subfloat[\label{subfig:trans}]{\hspace{.34\linewidth}}
\subfloat[\label{subfig:class}]{\hspace{.32\linewidth}}
\caption{Illustration of electron transport through a ring structure in \subref{subfig:quant} the quantum world, \subref{subfig:class} the classical or semiclassical world, and \subref{subfig:trans} in the transition regime between both worlds. \subref{subfig:quant} In the quantum world, plane electron waves interfere without any events. \subref{subfig:class} In the classical world, localized particles undergo numerous inelastic scattering events (red diamonds), which quickly erase the phase memory. \subref{subfig:trans} In the transition regime, wave packets interfere and undergo possible inelastic scattering events (halffilled diamond), which cause only a partial loss of coherence.}
\label{fig:trans}
\end{figure}

We analyze the mesoscopic transport of electrons by considering the event-type character of inelastic scattering, which initiates collapse processes \cite{on-the-sharp-of}. The idea behind this approach is presented in \figref{trans}. Whereas electron transport in the quantum regime is described by propagating extended waves and interference effects (\subfigref{quant}), it is characterized in the classical world by scattering events of particles (\subfigref{class}). These two transport regimes are exemplified by the Landauer-B\"uttiker formalism \cite{5392683,absen-of-backs-in} and the Drude-Sommerfeld model, described by the Boltzmann equation \cite{zur-elekt-der-metal,zur-elekt-der-metal-auf}. We focus here on the transition regime between classical and quantum transport and describe the charge carriers as coherent wave packets of finite size (\subfigref{trans}).
Existing formalisms assessing this regime are usually based on the Kubo formalism, which ensemble-averages the effects of collapses and dephasing processes \cite{stati-theor-of-irrev}. Such effects include the broadening of energy levels \cite{imry2002introduction} or ``washing out of states'' \cite{quant-oscil-and-the} and adding noise to wave functions' amplitudes and phases \cite{phase-uncer-and-loss,shot-noise-in-mesos,mesos-decoh-in-aharo}. These methods find a smooth crossover from the quantum regime to the classical world \cite{BEENAKKER19911,quant-trans-in-small} because decoherence
is modeled as a decay process phenomenologically.

In contrast, the quantum trajectory approach treats an inelastic event (a ``quantum jump'') as an individual event that does or does not take place. It breaks time-reversal symmetry by initiating a collapse of the wave function and occurs in real space at a distinct location and time \cite{on-the-sharp-of}. The quantum jumps can equivalently be described by an appropriate Lindblad master equation for the reduced density matrix of the observed system \cite{lindblad1976} which in our case consists of the electrons in a nanoscopic device. The spontaneous breaking of time-reversal invariance of the microscopic dynamics does not fit the assumptions underlying Onsager's reciprocity relations \cite{recip-relat-in-irrev,nonre-respo-from}. It is therefore not guaranteed that the transport must be reciprocal if quantum collapse processes are relevant.

This irreversibility is caused on a microscopic level by inelastic scattering events. It is therefore impossible to model it appropriately by a macroscopic procedure like tracing over unobserved degrees of freedom, as implemented by the methods mentioned above. Instead we shall use a description taking into account those events directly (see below). We exploit this intrinsic irreversibility to design novel electronic devices. These nonunitary quantum electronic devices work as follows. To start, the devices are nanostructured into noncentrosymmetric shapes. The spatial asymmetry induces a temporal asymmetry; the transit time $\tautr$ of a wave packet depends on whether the electron travels forward or backward through the device as shown in section \ref{sec-dyn}. This is because the transit time depends on the interference of the wave function due to the unitary dynamics. On the other hand, decoherence is caused by inelastic scattering events as characterized by the average time $\taui$ between such events. Therefore, the number of quantum jumps per passage of the electron through the device depends on $\taui/\tautr$ and is thus nonreciprocal. In one direction the electron transport is more coherent, in the other less so. The intrinsic irreversibility of the quantum mechanical decoherence process leads to nonreciprocity of electron transport.

Here, we demonstrate this device concept by modeling the electron transport in a set of noncentrosymmetric devices \cite{nonre-inter-for,lossl-curre-at-high,nonre-of-the-wavep}. These include devices with longitudinal asymmetry, devices with transverse asymmetry subject to a magnetic field, and two-path interferometers.
The devices are constructed from idealized model materials that are described by the single particle picture and by perfect screening.
The calculations were done by applying the Lindblad equation \cite{lindblad1976}, or using a direct, stochastic implementation of the wave-function collapse, by computing the quantum trajectory of the particle \cite{quant-measu-and,quant-stoch-proce-as,wavef-appro-to-dissi,carmichael2009open,quant-traje-and-open,molmer1993}. Both kinds of calculations cover the entire range from coherent quantum transport to classical diffusive transport.
As we work in the single particle picture, the Fermi statistics of the electron plays no role. For a recent proposal to implement Lindblad dynamics in the Meir-Wingreen transport formula, see \cite{gener-trans-formu,exact-descr-of-quant}.
Nonreciprocal transport of wave packets is achieved in all devices in a well defined window of scattering rates, yielding a peak of the nonreciprocity at a decoherence rate of ${\Gamma=2/\taui\approx1/\tautr}$ (see section \ref{sec-decoherence}). Crucially, in the limits of large and small scattering rates the transport is reciprocal. Therefore, this device concept illustrates the emergence of new functionality in electron transport right in the transition regime between quantum physics and classical physics. Having established the nonreciprocity for excited states (wave packets), we address in section \ref{sec-steady} the question whether the effect persists close to thermal equilibrium as described by a thermal density matrix. It turns out that the thermal equilibrium taken as a starting point of the state evolution is unstable against the collapse dynamics or decoherence. Novel time-independent states are established which show charge separation between the two leads, and, for closed systems, persistent currents.

\section{Nonreciprocal dynamics of wave packets for $\bm{\tau_i\gg\tau_\mathrm{tr}}$}
\label{sec-dyn}

\begin{figure}
\centering
\includegraphics[width=.8\linewidth]{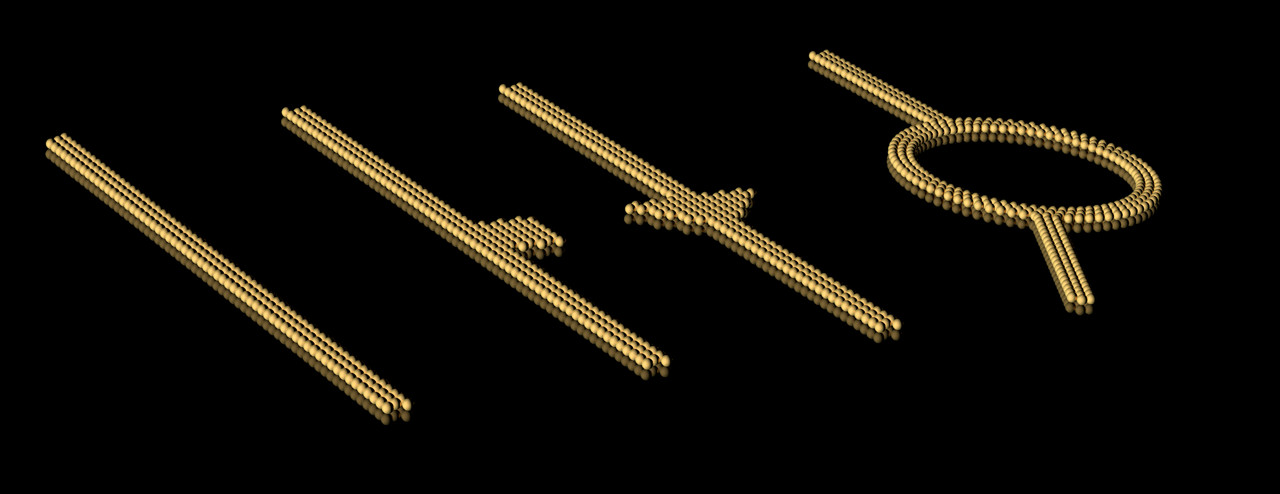}
\caption{\label{fig:sys-renders}Symmetric and asymmetric nanoscale conductors. The figure shows a symmetric line, conductors with a transverse asymmetry and a longitudinal asymmetry, and an example of an asymmetric Aharonov-Bohm ring (from left to right).}
\end{figure}

Figure~\ref{fig:sys-renders} shows the structure of the devices considered. The conductors connect two ports, $L$ and $R$, and are shaped asymmetrically perpendicular to or in the direction of the current flow (transversal and longitudinal asymmetry). We compare these conductors to Aharonov–Bohm rings \cite{signi-of-elect-poten, obser-of-frach-aharo} with a transversal asymmetry and to symmetric devices. To find the electron dynamics in the devices, we solve Schrödinger's equation for the given device geometries by exact diagonalization of the tight-binding Hamiltonian (see Appendix \ref{sec:numerics}).
Electron-electron interactions clearly modify the electron dynamics but, following the arguments used in Landauer's theory of ballistic transport \cite{5392683,absen-of-backs-in}, can be neglected to first approximation in these nanoscopic devices. Likewise, we do not consider additional elastic impurity scattering because this mechanism only adds the time-reversal invariant weak localization correction to the ballistic dynamics but does not affect the breaking of the time-reversal invariance which is the focus of the investigation.

The electron waves emitted by $L$ or $R$ into the one-dimensional conductor are described by Gaussian wave packets with momenta centered at ${k_F=\pi/(3a)}$, $a$ being the lattice constant. To consider transport across devices with a similar size as the electron wave packets, small, dispersive wave packets were chosen. The description in terms of wave packets, which forms the basis of the well-established semiclassical description of electron dynamics \cite{solid-state-physi}, is central to our calculations, because Bloch-waves lack time dependence. Since we use the exact eigenfunctions of the single-particle Hamiltonian to compute the time evolution of the states, the results in this section are obtained in the unitary quantum regime.

\begin{figure*}
\centering
\subfloat[\label{subfig:sys-lrat}]{
\begin{minipage}[t]{.49\linewidth}
  \centering
  \includegraphics[scale=1]{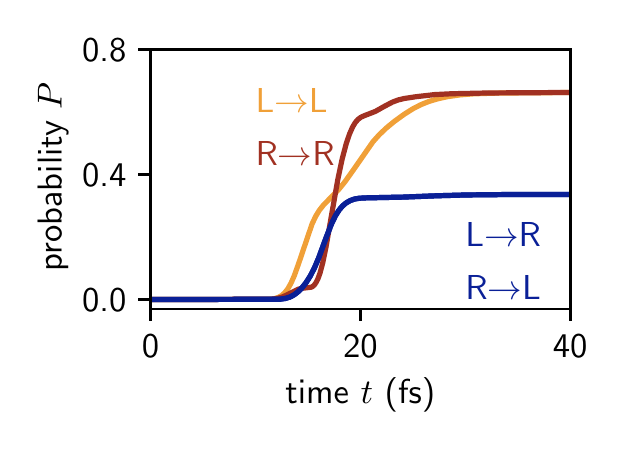}\\
  \includegraphics[scale=.2]{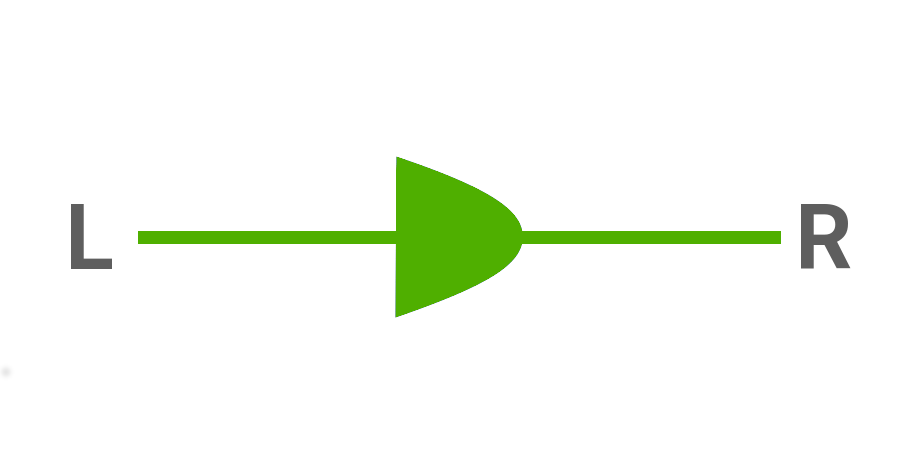}
\end{minipage}}
\subfloat[\label{subfig:sys-trat}]{
\begin{minipage}[t]{.49\linewidth}
  \centering
  \includegraphics[scale=1]{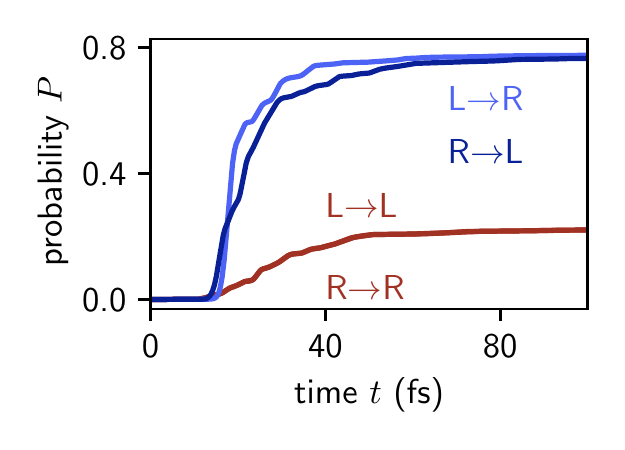}\\
  \includegraphics[scale=.2]{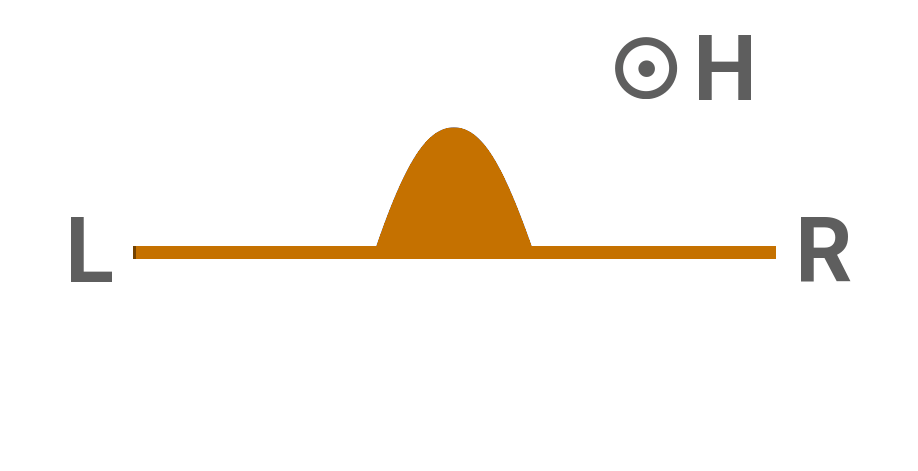}
\end{minipage}}
\\
\subfloat[\label{subfig:sys-amzi}]{
\begin{minipage}[t]{.49\linewidth}
  \centering
  \includegraphics[scale=1]{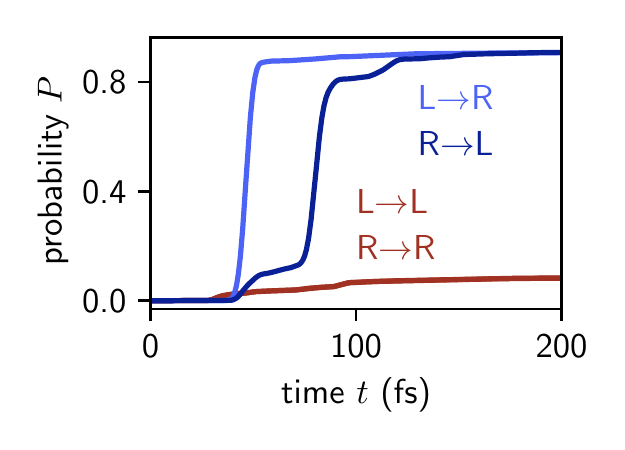}\\
  \includegraphics[scale=.2]{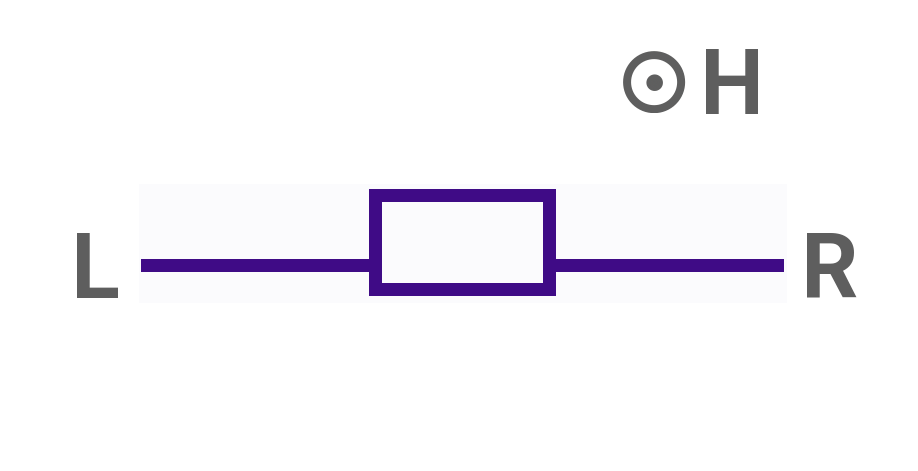}
\end{minipage}}
\subfloat[\label{subfig:sys-sdot}]{
\begin{minipage}[t]{.49\linewidth}
  \centering
  \includegraphics[scale=1]{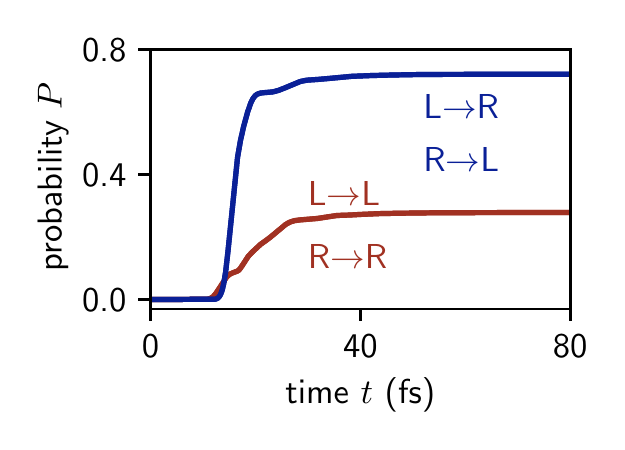}\\
  \includegraphics[scale=.2]{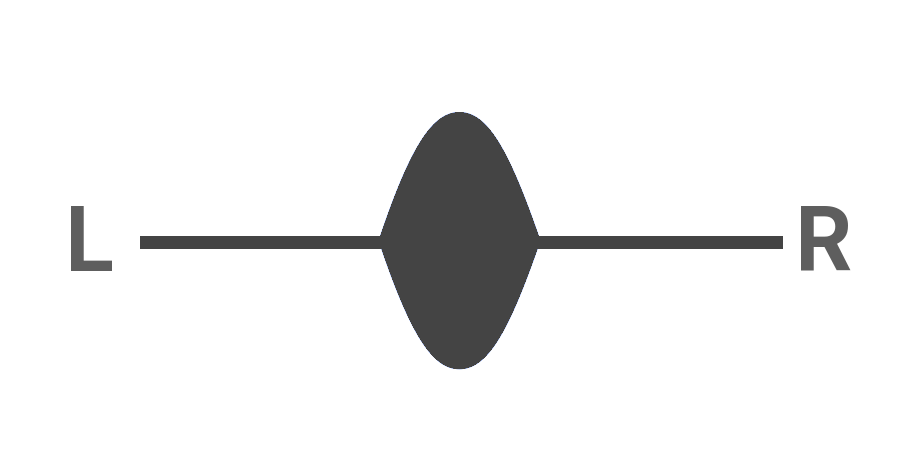}
\end{minipage}}
\caption{Layouts and time resolved transmission and reflection probabilities of unitary conductors with \subref{subfig:sys-lrat} a longitudinal asymmetry, \subref{subfig:sys-trat} a transversal asymmetry, \subref{subfig:sys-amzi} an asymmetric Aharonov–Bohm loop, and \subref{subfig:sys-sdot} a symmetric device.
The devices
\subref{subfig:sys-trat} and \subref{subfig:sys-amzi} with areas ${28\,a^2}$ and ${398\,a^2}$ are penetrated by magnetic fluxes $\Phi=\SI{1.4}{h/e}$
and \SI{0.4}{h/e}, respectively, where ${a=\SI{.3}{\nm}}$ is the unit length of the tight-binding lattice and $h/e$ is the magnetic flux quantum.
The structures connect two ports $L$ and $R$. In the plots, the transmission probabilities $\Pmn$ for electrons emitted as Gaussian wave packets (${k_0=\pi/3a}$, ${\sigma_x=\SI{0.9}{\nm}}$ for (\subref*{subfig:sys-lrat},\subref*{subfig:sys-trat},\subref*{subfig:sys-sdot}) and \SI{24}{\nm} for \subref{subfig:sys-amzi}) by port $m$ to reach port $n$ calculated as a function of time after emission are presented for the scattering-free, unitary transport.
The devices' lengths are \subref{subfig:sys-lrat} \SI{1.8}{\nm}, \subref{subfig:sys-trat} \SI{3.0}{\nm},
\subref{subfig:sys-amzi} \SI{12}{\nm} and \subref{subfig:sys-sdot} \SI{2.4}{\nm}.}
\label{fig:dwell}
\end{figure*}

Figure \ref{fig:dwell} displays the probability that the electron has reached one of the ports of the conductor with the transverse asymmetry as a function of time $t$ after the electron emission. As demanded by the unitarity of the scattering matrix, these probabilities are reciprocal in the long-time limit \cite{imry2002introduction, BEENAKKER19911}. In contrast, the time evolution does depend on the travel direction \cite{nonre-inter-for,lossl-curre-at-high,nonre-of-the-wavep}. The left-right symmetry is broken because the phase shifts of the individual plane waves are nonreciprocal. As the phase shifts are also $k$-dependent, they influence the temporal behavior of the wave packet, which comprises many plane waves with different $k$-values.
This leads to $\tauLR \neq \tauRL$, where $\tauLR$ denotes the time spent in the device by a wave packet coming from port $L$ before it leaves through port $R$ (see also videos~\ref{vid:trat}, \ref{vid:amzi}).

We find that a nonreciprocal temporal dependence of particle transport is a generic property of many quantum devices with appropriately broken symmetries. \subfigref{sys-lrat} shows a device that is symmetric in the transversal direction, but asymmetric in the longitudinal direction. In these devices, the reflected electrons follow a nonreciprocal temporal dependence even without an applied magnetic field. The nonreciprocal temporal dependence has been predicted for asymmetric Rashba rings and for Aharonov–Bohm rings biased with magnetic fields \cite{nonre-inter-for, lossl-curre-at-high} (\subfigref{sys-amzi}, videos \ref{subvid:amzi-l} and \ref{subvid:amzi-r}). Note that the reflection probabilities $\PLL$, $\PRR$ are equal for the systems in
\subfigref{sys-trat} and \subfigref{sys-amzi}, while the transmission probabilities $\PLR$, $\PRL$ are equal in the system of \subfigref{sys-lrat} without magnetic field. This feature reflects the symmetries of the scattering matrix under spatial inversion and time-reversal \cite{nonre-of-the-wavep}. In contrast, conductors without adequately broken symmetries (\subfigref{sys-sdot}) show reciprocal dynamics.

\section{Nonreciprocal behavior of wave packets for any $\bm{\tau_i}$}
\label{sec-decoherence}

In order to calculate the transport in the transition regime between unitary quantum physics and classical physics, we now add decoherence to the dynamics of the wave packets. This decoherence is commonly characterized by the phase-breaking time $\taui$. We expect the physics of the transition regime to emerge when the defect density causes $\taui$ to be of the order of the average electron transmission time ${(\tauLR+\tauRL)/2}$. As shown, the electron transmission time is not necessarily equal for both transmission directions. In those cases, electrons that pass the device in the slower direction suffer stronger decoherence than those that pass in the faster one. The transmission probabilities for both directions, however, are guaranteed to be equal only if the degree of phase-breaking is equal in both directions. Therefore, the transmission probabilities are expected to be possibly nonreciprocal in devices with nonreciprocal decoherence.

%
The interaction with the environment is mediated by one or more sites of the tight-binding lattice that we use to model the device, see Appendix \ref{sec:numerics}. Apart from the localized nature of the interaction, we introduce no constraint on the nature of the decoherence or collapse processes, which may be caused e.g.\ by local lattice distortions (local phonons) or the internal degrees of freedom of an impurity giving rise to a localized resonant level. We are interested in a description of the electron dynamics, i.e. the time evolution of the reduced density matrix $\rho_e(t)$ of the (single) electron. This reduced matrix is obtained from the full density matrix by tracing over the unobserved degrees of freedom. The general form of the resulting time evolution is a time-local master equation for $\rho_e(t)$ that preserves complete positivity \cite{lindblad1976}:
\begin{align}
  &\frac{\diff \rho_e}{\diff t} = -\frac{i}{\hbar}[H_0,\rho_e] 
  +\sum_{\bm{r}}\gamma_{\bm{r}}\cD[\hat{P}_{\bm{r}}](\rho_e)
      .
  \label{eqn:lindblad1}
\end{align}
Here, $H_0$ is the single particle Hamiltonian of the system without the impurities.
The Lindblad dissipators $\cD[\hat{P}_{\bm{r}}](\cdot)$ depend on the index $\bm{r}$ of the site at which an impurity and the concomitant electronic level are located. They are generally defined as
\begin{equation}
 \cD[\hat{A}](\rho_e)= \hat{A}\rho_e\hat{A}^\dagger -
      \frac{1}{2}(\hat{A}^\dagger\hat{A}\rho_e +\rho_e\hat{A}^\dagger\hat{A}),
\label{eqn:dissipator}
\end{equation}
where $\hat{A}$ denotes the so-called jump operators.
For devices A, B and D we choose hermitean jump operators $\hat{P}_{\bm{r}}$ that project the electron wave function onto the site $\bm{r}$, i.e. $\hat{P}_{\bm{r}}=|\bm{r}\rangle\langle\bm{r}|$, with $\hat{P}_{\bm{r}}=\hat{P}_{\bm{r}}^\dagger=\hat{P}_{\bm{r}}^2$.
This mechanism corresponds to a measurement of the occupation of site $\bm{r}$ without read-out. For device C we chose a projection onto the full subspace spanned by the sites of one interferometer arm. In both cases the localization erases the information about the momentum and phase of the electron trapped at the site $\bm{r}$.
All jump operators are multiplied with the $\bm{r}$-independent rate $\Gamma=\gamma_{\bm{r}}=2/\taui$, where $\taui$ is the mean time between two phase-breaking scattering events.
After release from the trap, the electron state evolves again according to the Schr\"odinger equation given by $H_0$.

\begin{figure*}
  \centering
  \subfloat[\label{subfig:peak-lrat}]{
    \includegraphics[scale=1]{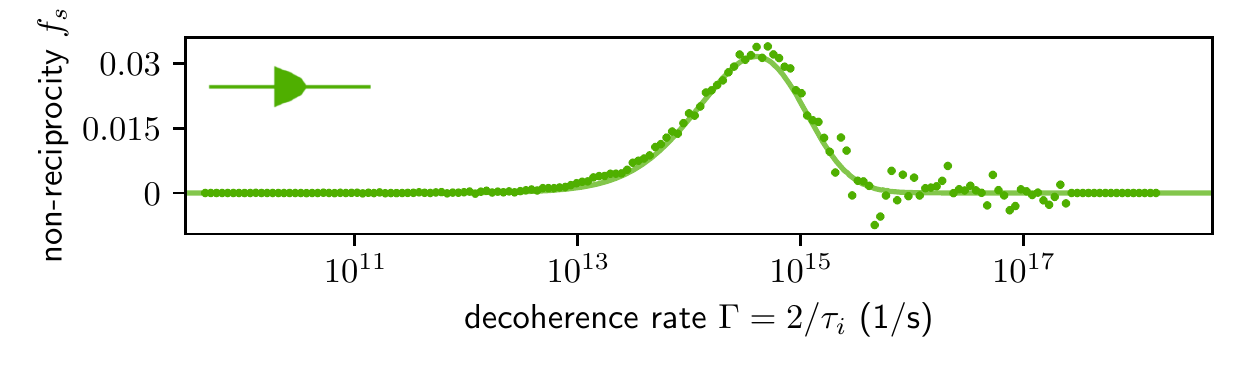}}\\
  \subfloat[\label{subfig:peak-rest}]{
    \includegraphics[scale=1]{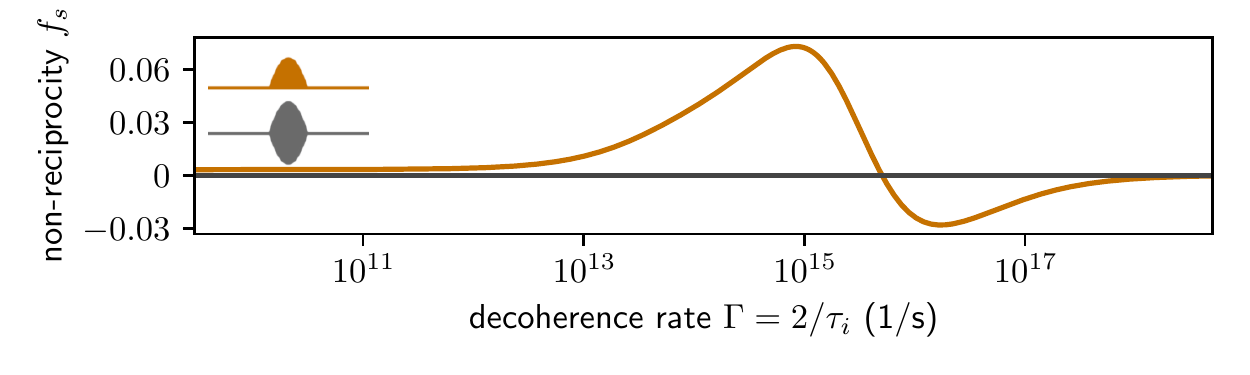}}\\
  \subfloat[\label{subfig:peak-amzi}]{
    \includegraphics[scale=1]{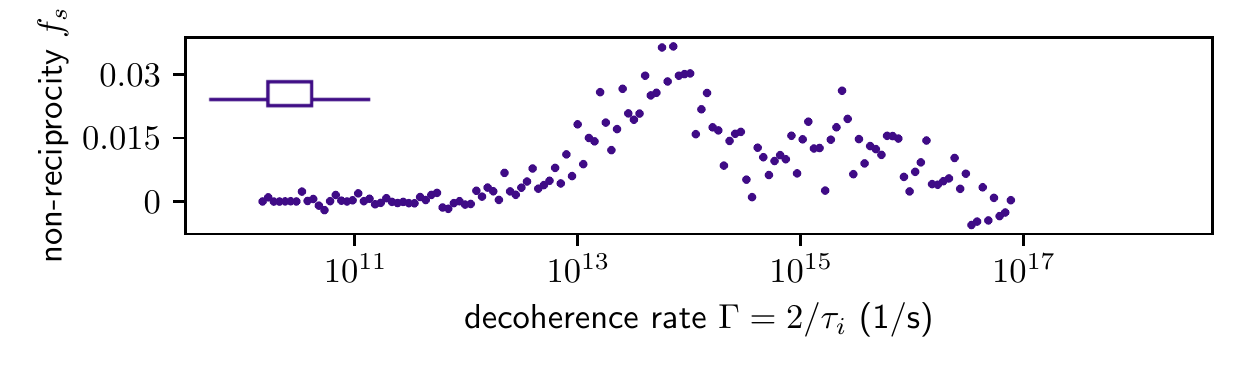}}
  \caption{Nonreciprocity of the devices introduced in Fig.~\ref{fig:dwell} calculated with the Lindblad formalism (continuous lines) and the Monte-Carlo wave function method (dotted). \subref{subfig:peak-lrat} Nonreciprocal transport through the device A (Fig.~\ref{subfig:sys-lrat}) occurs for ${\invtcnon<\Gamma<\invtcdiff}$. As anticipated, both methods yield coincident results. The Monte-Carlo data was obtained from \lratValidTrajectories{} realizations of the time evolution.
  \subref{subfig:peak-rest} Nonreciprocal transport of device B (orange, Fig.~\ref{subfig:sys-trat}) in the transition regime and return to reciprocal behavior for large and small $\Gamma$. The transport through the symmetric device D (gray, Fig.~\ref{subfig:sys-sdot}) is fully reciprocal, however.
  \subref{subfig:peak-amzi} Nonreciprocal transmission of device C (Fig.~\ref{subfig:sys-amzi}) as obtained with the Monte-Carlo wave function method.}
  \label{fig:nonreci}
\end{figure*}

To quantify the effects of the decoherence on the transport, we prepare the electron at time $t=0$ as a Gaussian wave packet which enters the system either at port $L$ or port $R$. We integrate the time-dependent local probability current at site $\bm{r}_{L,R}$ in the left or right port until time $t_{\text{fin}}$ to obtain transmission and reflection probabilities (see Appendix \ref{sec:currents}). Reciprocal transport demands $\PLR=\PRL$ for sufficiently large $t_{\text{fin}}$. Figure~\ref{fig:nonreci}
shows the nonreciprocity $f_s=\PLR-\PRL$ as function of $\Gamma$ computed with \eqnref{lindblad1} and a set of localization centers at positions $\bm{r}_1,\ldots\bm{r}_n$ within the devices (see Appendix \ref{sec:numerics}). For the ballistic regime, i.e. for small $\Gamma$, $f_s$ indeed vanishes.

However, sorting occurs in the asymmetric devices A, B and C in a specific window of $\Gamma$. In this window, electrons that are emitted by $L$ and $R$ reach port $R$ with a higher probability than port $L$. The window for nonreciprocal transport matches the inverse device transit time which is a function of the properties of the chosen wave packet. Furthermore, we observe a restoration of reciprocity at high inelastic scattering rates $\Gamma$, corresponding to the crossover to classical diffusive transport. Notably, the sorting is completely absent for the symmetric device D.

The physics behind the nonreciprocal transport happening in a well-defined window of $\Gamma$ may be intuitively understood by unraveling the Lindblad equation \eqnref{lindblad1} as a stochastic evolution of the wave function. Following this concept, we consider events at randomly chosen times $t_1\ldots t_n$ with a mean spacing ${\langle t_{j+1}-t_{j}\rangle=\taui=2/\Gamma}$. As before, the rate $1/\taui$ of this Poisson process is proportional to the coupling between electrons and the defect or phonon systems. Between the ``collapse times'' $t_{j}$ and $t_{j+1}$ the wave function evolves unitarily with the Hamiltonian $H_0$. At $t_{j}$ and $t_{j+1}$ the wave function changes according to the corresponding jump operator $\hat{P}_{\bm{r}}$.
Two outcomes can be distinguished: In the first case, the wave function is projected onto an eigenstate of the operator $\hat{P}_{\bm{r}}$, it becomes $|\bm{r}\rangle$  with probability $p_+=|\langle\psi|\bm{r}\rangle|^2$ as given by the Born rule \cite{von2013mathematische}. In the second case the ``measurement result'' is negative: The state is projected onto the orthogonal complement of $|\bm{r}\rangle$ and changes to $|\bm{r}^\perp\rangle$ with probability $p_-=1-p_+$.
For $t>t_{j+1}$, the wave packet evolves again unitarily until it undergoes a second random inelastic scattering event at $t_{j+2}$ or leaves the system via the two ports (see video~\ref{vid:amzi-closed}).

Due to the stochastic time-evolution, the transmission probability $\PLR$ is itself a random variable. To obtain its distribution function $n_{\text{L}\rightarrow\text{R}}=n(\PLR)$, we consider an ensemble of many stochastic quantum trajectories. Figure~\ref{subfig:peak-lrat} shows the average $\langle\PLR\rangle-\langle\PRL\rangle$ for \lratValidTrajectories such evolutions for device A. As shown in the figure, the results obtained by this stochastic implementation of the wave-function collapse are quantitatively consistent with the calculation based on the Lindblad-formalism.

\begin{figure}
  \centering
  \subfloat[\label{subfig:avg-colormap}]{
    \includegraphics[scale=.7]{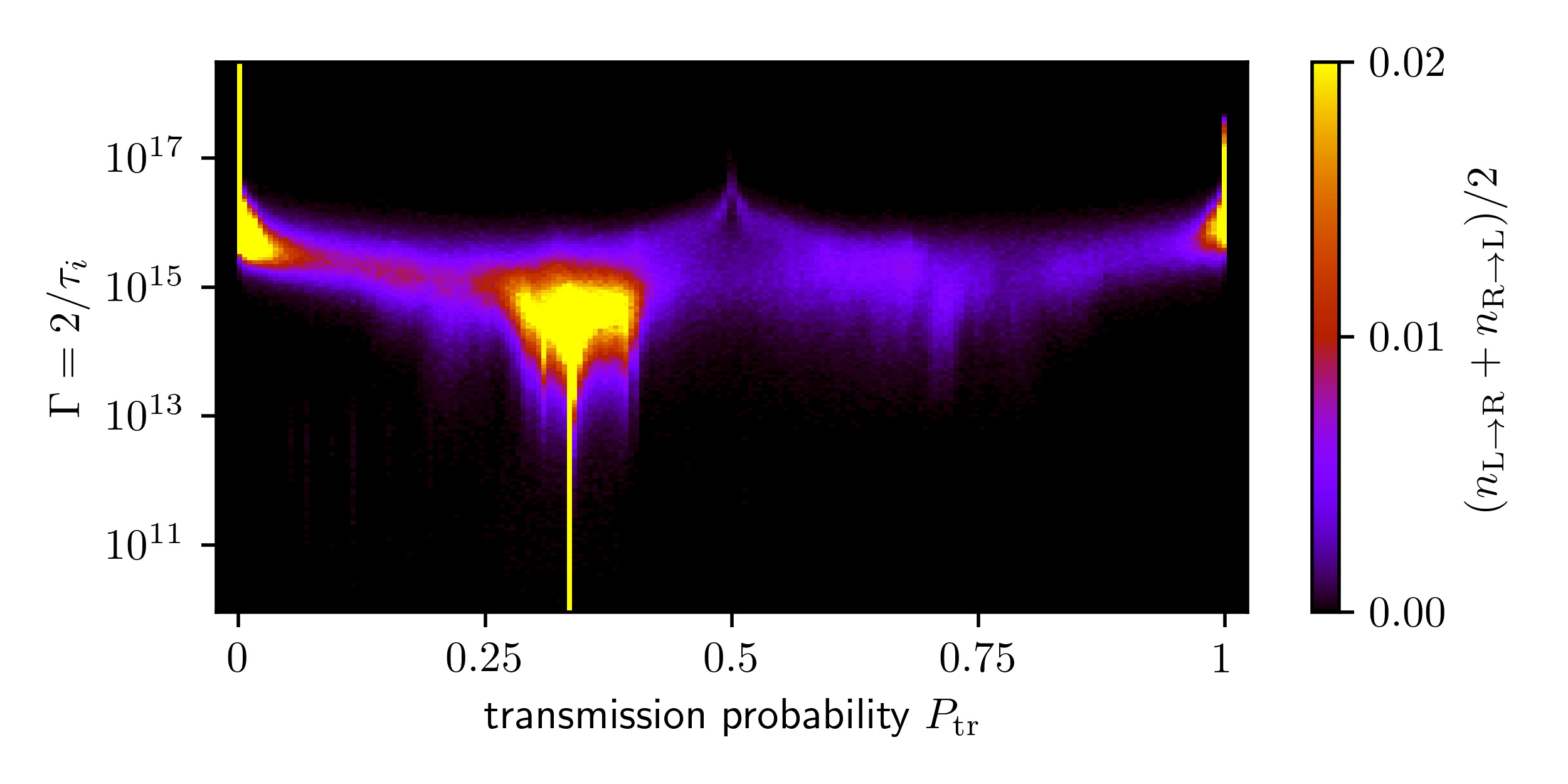}}\\
  \subfloat[\label{subfig:diff-colormap}]{
    \includegraphics[scale=.7]{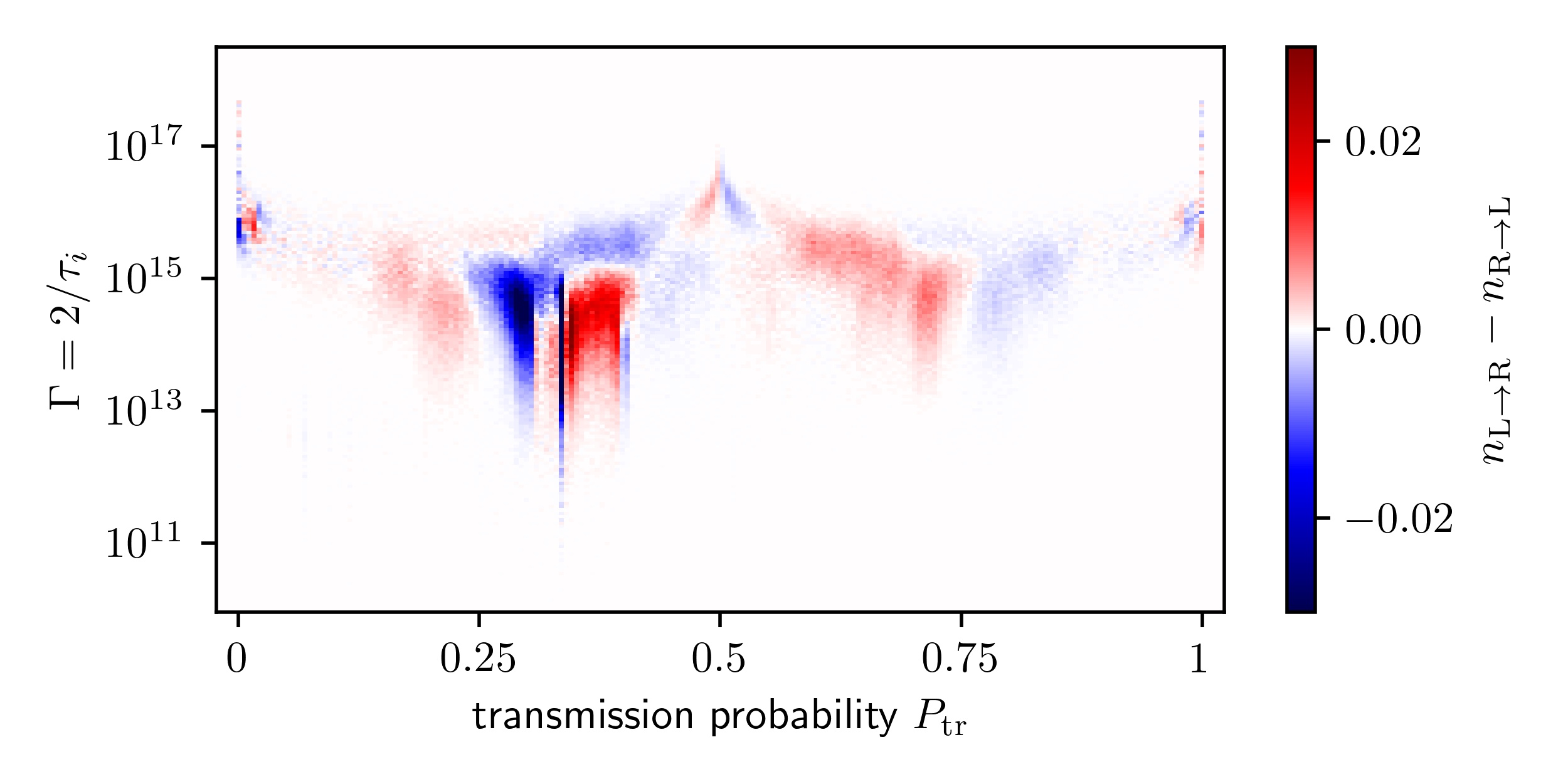}}
  \caption{Colormaps showing the transmission probability distribution as a function of $\Ptr$ and decoherence rate $\Gamma$ for the device A. \subref{subfig:avg-colormap} The direction-independent mean of the distributions for $L\rightarrow R$ and $R\rightarrow L$ directions illustrate the different transport regimes. \subref{subfig:diff-colormap} The difference of the  distributions for the two directions shows the direction dependent effect of decoherence, which leads to nonreciprocal dynamics on long time scales in the transition regime.}
  \label{fig:colormaps}
\end{figure}

We use the stochastic approach to illustrate the transition between the different regimes of electron transport. Figure~\ref{subfig:avg-colormap} shows the distribution $n(\Ptr,\Gamma)$ of the transmission probability $\Ptr$ of device A as a function of $\Gamma=2/\taui$ averaged over the two travel directions. For ${\Gamma\lesssim\invtcnon}$, in the unitary quantum regime, the probability for $\Ptr$ is peaked at $0.3$, indicating that individual electrons leave the system in a state that is a superposition of \SI{70}{\percent} being located in the port of origin and \SI{30}{\percent} in the other port.
For ${\invtcnon\lesssim\Gamma\lesssim\invtcdiff}$ --- in the transition regime --- the distribution varies rapidly with changing $\Gamma$ and superpositions of any composition are found. For ${\invtcdiff\lesssim\Gamma\lesssim\invtczeno}$, in the classical regime, the distribution is peaked at
$\Ptr=0$ and $\Ptr=1$. Each individual electron is either reflected or transmitted by the device. States with a coherent superposition are no longer possible. Finally, for ${\Gamma\gtrsim\invtczeno}$, the probability has a single peak at $\Ptr=0$. This is due to the quantum Zeno effect \cite{misra1977}: the scattering rate is so high that electrons cannot pass through the device. Figure~\ref{subfig:diff-colormap} shows the difference of the distributions belonging to the two directions. We draw attention to three features of the graph. First, the largest difference is observed at $\Ptr=0.3$, the most probable transmission in the case without scattering. This implies that electrons traveling in one of the directions are indeed more frequently scattered inelastically than electrons traveling in the opposite direction. Second, positive differences are found at higher values of $\Ptr$ than negative differences. This implies nonreciprocity of the average transmission. Finally, the nonreciprocity is easily observed when approaching the classical regime: the peak at $\Ptr=0$ is negative and the peak at $\Ptr=1$ is positive. Therefore, more electrons are transmitted when originating from $L$ and more electrons are reflected when originating from $R$. The stochastic time evolution of single electrons provides an understanding of the dynamics in the transition regime and confirms the reasoning that the observed nonreciprocity is caused by direction dependent decoherence.

\section{The charge-separated steady state}
\label{sec-steady}

In the previous two sections we have shown that nonreciprocal transport is present when electrons are described as wave packets.
How are these results to be interpreted in the larger framework of transport theory which refers to the states in the vicinity of thermal equilibrium, characterized by a well-defined constant temperature on the nanoscopic scale?
In equilibrium, electron wave packets may occur as fluctuations, e.g.\ by the thermally excited release of electrons from trapping sites. Indeed, the success of the semi-classical theory of electron transport shows unambiguously that physically relevant thermal excitations from the Fermi sea are wave packets of finite size.
According to Onsager's reciprocity relations, such fluctuations always relax back to the equilibrium state, which is the starting point of both the semi-classical and quantum theories of linear transport
\cite{5392683,absen-of-backs-in,zur-elekt-der-metal,zur-elekt-der-metal-auf, stati-theor-of-irrev,quant-oscil-and-the,phase-uncer-and-loss,shot-noise-in-mesos, mesos-decoh-in-aharo,BEENAKKER19911,quant-trans-in-small}.

\begin{figure}
  \centering
  \begin{tikzpicture}[scale=0.8]
    \draw[line width=2, tabblue] (-4,0) -- (4,0);
    \foreach \i in {1,...,9} {
      \draw[tabblue, line width=2, fill=tabblue!10!white] (\i-5,0) circle (.3);
      \node at (\i-5,-.9) {\ket{\i}};
    }
    \draw[line width=0, fill=tabred] (-.15,-.5) -- (.15,-.5) -- (0,-.3) -- (-.15,-.5);
    \draw[line width=0, fill=tabred] (-.15,.5) -- (.15,.5) -- (0,.3) -- (-.15,.5);
    \draw[tabblue, line width=2, fill=tabblue!30!white] (1,0) circle (.3);
    \draw[tabblue, line width=2, fill=tabblue!60!white] (0,0) circle (.3);
    \draw[tabblue, line width=2, fill=tabblue!90!white] (-1,0) circle (.3);
    \draw [gray,decorate,line width=1.5,
           decoration={brace,amplitude=8,aspect=.5}]
        (-4.3,.7) -- (-1.7,.7) node[pos=.5,above=.8em] {$Q_\mathrm{L}$};
    \draw [gray,decorate,line width=1.5,
           decoration={brace,amplitude=8,aspect=.5}]
        (-1.3,.7) -- (1.3,.7) node[pos=.5,above=.8em] {$Q_\mathrm{C}$};
    \draw [gray,decorate,line width=1.5,
           decoration={brace,amplitude=8,aspect=.5}]
        (1.7,.7) -- (4.3,.7) node[pos=.5,above=.8em] {$Q_\mathrm{R}$};
  \end{tikzpicture}
  \caption{Sketch of the tight-binding chain used for most of the calculations in Section \ref{sec-steady}. The blue circles represent the tight-binding sites.
  The fill color of the sites sketches the local value of the electrostatic potential. The darker the shade, the higher is the potential at that site.
  The sites \ket{4}, \ket{5} and \ket{6} implement an asymmetric potential barrier. The central site \ket{5} highlighted by the red arrows is coupled to a trap, which may absorb the particle and release it again as a wave packet. The observables ${Q_\mathrm{L}, Q_\mathrm{R}}$ and ${Q_\mathrm{C}}$ are used to measure the charge density in the left, right and central region of the open system shown, respectively.}
  \label{fig:lindblad-chain-sys}
\end{figure}
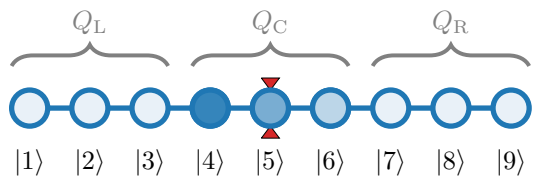

For the devices presented here, however, these fluctuations do not necessarily relax back into the standard thermal equilibrium state. Wave packets that occur as fluctuations around the thermal equilibrium may be sorted by the devices as discussed in Sections \ref{sec-dyn} and \ref{sec-decoherence} and therefore lead to a charge accumulation at one side of the device. To demonstrate this numerically, we employ a minimized version of the models used in the previous sections. This system, sketched in \figref{lindblad-chain-sys}, is given by a chain of 9 sites with the site basis vectors ${\ket{1}, \ket{2}\dots\ket{9}}$.
A ramp-shaped electric potential is applied to sites \ket{4}, \ket{5} and \ket{6} to break the symmetry. The potential at the remaining sites is zero (see Appendix \ref{sec:minisys}).

\begin{figure*}
  \centering
  \subfloat[\label{subfig:lindblad-init-therm}]{
    \includegraphics[scale=1]{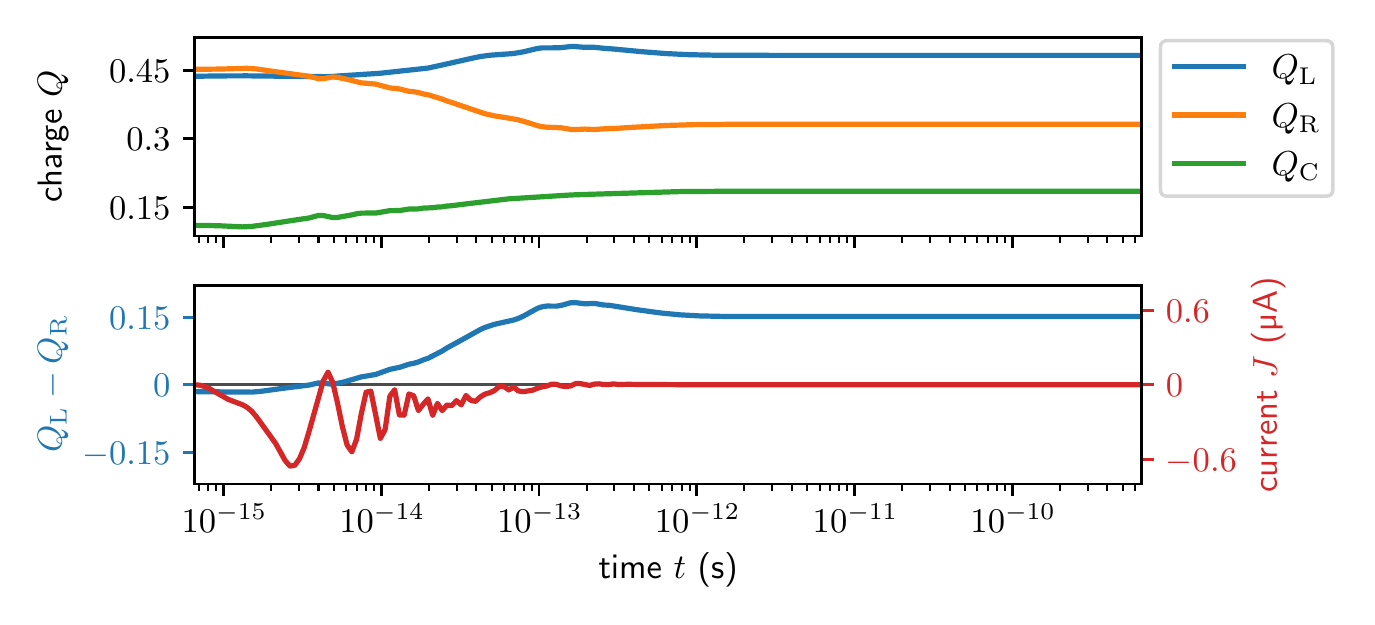}
  }\\
  \subfloat[\label{subfig:lindblad-init-therm-currents}]{
    \hspace{.5cm}
    \includegraphics[scale=1]{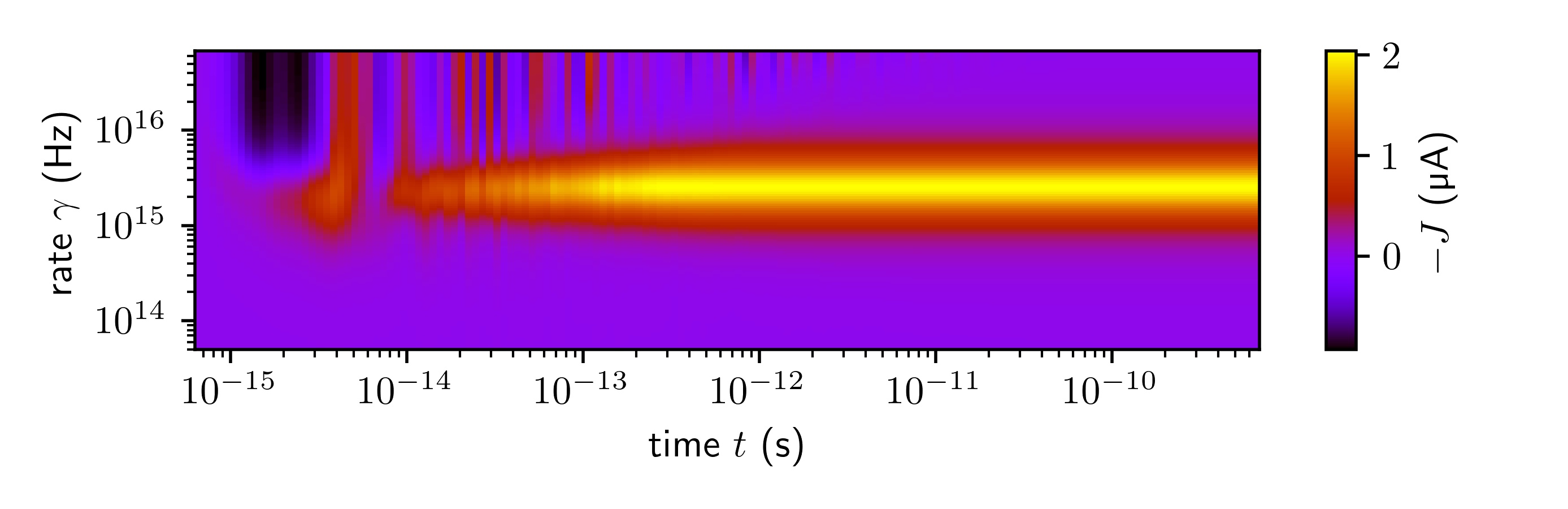}
  }\\
  \subfloat[\label{subfig:lindblad-init-therm-peak}]{
    \hspace{.5cm}
    \includegraphics[scale=1]{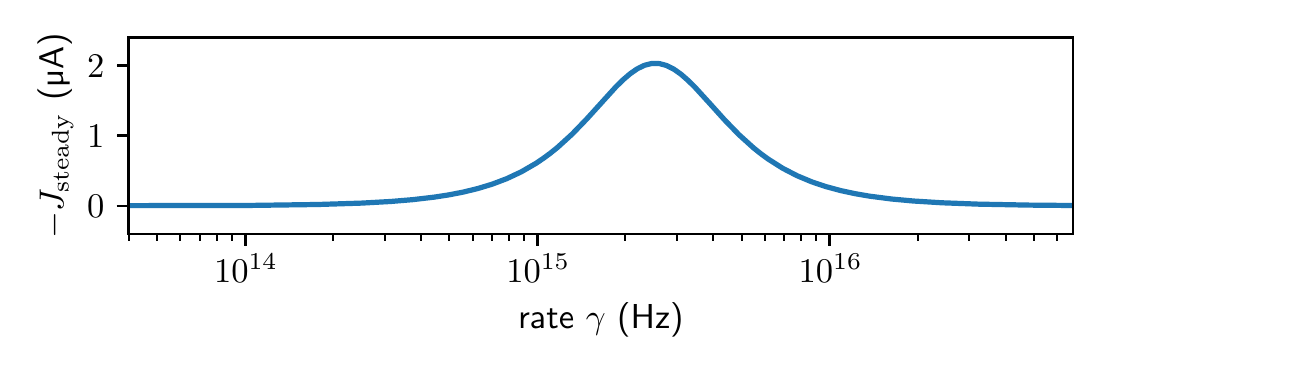}
  }
  \caption{
  \subref{subfig:lindblad-init-therm}~Plots showing the time evolution of charges and the total current of the open 9-site chain. The initial state is given by a thermal density matrix with ${kT=\SI{1}{\eV}}$. The thermal density matrix is not a steady state of the dynamics and the system transitions within ${\approx10^{-12}\,\si{s}}$ to the new charge-separated steady state.
  The initial charge separation is caused by the asymmetric electrostatic potential. This charge difference is reversed through the nonreciprocal collapse dynamics: the electrochemical potential is no longer the same in both ports.
  \subref{subfig:lindblad-init-therm-currents}~The plot of the total current $J$ as a function of time $t$ and wave-packet generation rate $\gamma$ for the closed (shorted) device shows that a finite steady-state current is achieved for ${\gamma\approx\SI{5e14}{\Hz}\dots\SI{2e16}{Hz}}$.
   The time scale of the transition to the new steady state is again ${\approx10^{-12}\,\si{s}}$.
  \subref{subfig:lindblad-init-therm-peak}~Steady-state current $J_\mathrm{steady}$ plotted as function of the scattering rate $\gamma$. The current develops a peak at $\gamma\approx\SI{2e15}{\Hz}$. This peak is the counterpart of the peak shown by the nonreciprocity of the wave-packet transmission probability as function of $\Gamma$ (\subfigref{peak-lrat}).
  }
  \label{fig:lindblad-thermo}
\end{figure*}

Because the existence of wave packets is crucial, we modify the jump operators in \eqnref{lindblad1} to now describe the capture of the electron at site $\bm{r}$ and the subsequent release from this site into the state $|\psi_{\bm{r}}(\bm{p})\rangle$ which is a wave packet. The wave packet is generated at ${\bm{r}}$ with momentum ${\bm{p}}$ which is determined by the recoil of the trap which now carries the momentum $-\bm{p}$.
For the momentum $\bm{p}$ the value ${\pi/3a}$ is used because it is in the range of momenta for which relevant interference effects take place in the device. The following results do not depend on the choice of ${\bm{p}}$
in a qualitative manner.
The corresponding jump operators $\hat{A}_{\bm{r}}$ are non-hermitean,
$\hat{A}_{\bm{r}}=|\psi_{\bm{r}}(\bm{p})\rangle\langle \bm{r}|$ with the Lindblad equation
\begin{equation}
  \frac{\diff \rho_e}{\diff t} = -\frac{i}{\hbar}[H_0,\rho_e]
  +\sum_{\bm{r}}\gamma_{\bm{r}}\cD[\hat{A}_{\bm{r}}](\rho_e).
  \label{eqn:lindblad2}
\end{equation}
\subfigref{lindblad-init-therm} shows the time-dependent probability for the electron being in the left, right or central part of the chain, as derived from the dynamics of \eqnref{lindblad2} with wave packets being generated at site \ket{5} with the rate $\gamma=\SI{1.5e15}{Hz}$. The initial state is given by the thermal density matrix $\rho_e(0)=\sum_i|E_i\rangle\langle E_i|e^{-E_i/kT}$, where $|E_i\rangle$ are the eigenstates of $H_0$ with energy $E_i$
and ${kT=\SI{1}{\eV}}$. As shown by \figref{lindblad-thermo}, this state is not stable, but evolves into a steady state with a lower entropy and partial charge separation between the two ports. The chain spontaneously develops a charge imbalance with net current zero.
The $I$-$V$ characteristic contains therefore a new, constant term,
\begin{equation}
  V(I) = V_0 +RI + \mathcal{O}(I^2).
  \label{eqn:VIchar}
\end{equation}
The central assumption underlying Onsager's reciprocity relations is therefore not satisfied in the case discussed. Furthermore, this $I$-$V$ characteristic suggests a net current to flow in case the chain is bent into a ring, with the left and right chain ends being shorted. Figs.~\ref{subfig:lindblad-init-therm-currents} and \ref{subfig:lindblad-init-therm-peak}, showing the results of the corresponding calculation, reveal that a circulating persistent current is indeed generated
for ${\gamma\approx\SI{5e14}{\Hz}\dots\SI{2e16}{Hz}}$, corresponding to the transition regime.
This persistent current induces a corresponding magnetic moment.
Notably, this steady-state current vanishes for very small and large rates $\gamma$. This behavior is analogous to the behavior of the wave-packet transmission probabilities, which are nonreciprocal only in a well-defined window of the decoherence rate.

To investigate whether the asymmetric device geometry is responsible for the charge separation and the circulating current, the state evolution of a chain and a ring with symmetric potential barriers were calculated. As expected, these displayed no charge separation (\figref{lindblad-thermo-sdot}).
With hermitean jump operators (\eqnref{lindblad1}), also no charge separation is found in the long time limit, even when starting from a non-equilibrium density matrix (\figref{lindblad-steady-pack}).
This underlines the importance of the existence of wave packets with non-zero momentum for the charge-separated steady state.
To explore whether the unconventional steady state generated by \eqnref{lindblad2} is destroyed by other inelastic processes that may exist and drive the system towards standard equilibrium, we have also added such processes to the master equation and found that the novel steady state persists also in this case (see Appendix \ref{sec:minisys}).

\section{Discussion and Outlook}

The transition between the quantum and classical worlds is of intense interest. It harbors fundamental questions concerning the appropriate description of decoherence and the measurement process. The devices we have discussed operate precisely at the border between these two worlds, because they utilize a small number of random phase-breaking events that interrupt the otherwise unitary evolution of wave packets. Our work shows that in the unitary regime, electrons flow through devices with nonreciprocal velocities if the devices are shaped with appropriate asymmetries.
Our results have been obtained by using several assumptions and are only valid in those cases in which these assumptions apply. In particular, we have considered idealized model systems following a strict single particle picture with perfect screening and without disorder. For those real materials that may be influenced by the phenomena described, e.g. lightly doped semiconductors,
nothing more than nanostructuring a film is required in order to achieve rectification as described by \eqnref{VIchar}.
The optical analog of such devices is presented in \cite{fermi-golde-rule-and}.
The effects presented differ from the nonreciprocal behavior of standard diodes \cite{can-the-recti-becom}, quantum rings \cite{magne-field-symme}, quantum dots \cite{exper-inves-of-the}, chiral structures \cite{elect-magne-aniso},
Weyl semimetals \cite{chira-anoma-and}, noncentrosymmetric superconductors \cite{highl-cryst-2d-super}, and multiferroics \cite{obser-of-ferro-domai}. In those cases, the nonreciprocity is achieved by nonlinear, higher-order processes; the voltage for $I\rightarrow 0$ vanishes, $V_0=0$. In those cases, also no unconventional steady state exists besides the standard thermal equilibrium.

In conclusion, we have presented a device concept in which nonreciprocal matter transport emerges when the inverse decoherence rate is of the order of the characteristic time for unitary transport through the device. This situation exists exactly in the transition regime between quantum physics and classical physics. The nonreciprocal matter transport is expected to occur not only in top-down patterned devices but also in molecules with appropriate asymmetric structures and in crystals with suitable lattices. The phenomena found are explorable by experiments on mesoscopic electronic or photonic devices. The described mechanism underlying nonreciprocal, directed dynamics could even be responsible for the proper operation of biomolecules and thus for living systems.

\section{Acknowledgement}

The authors gratefully acknowledge useful discussions with A. Alavi, A. Brataas, T. Kopp, P. Schneeweiss, and support by L. Pavka, T. Whittles, and the computer service group of MPI-FKF. J. Mannhart acknowledges support by the Center for Integrated Quantum Science and Technology (IQST). The numerical calculations were performed using the Kwant \cite{kwant-a-softw-packa} and QuTIP \cite{qutip-an-opens-pytho,qutip-2-a-pytho} Python packages.

{\color{white} basd asd asd asd asdas dasd asd asda sdasd asd asd asd}
{\color{white} basd asd asd asd asdas dasd asd asda sdasd asd asd asd}
\bibliography{main}

\appendix

\cleardoublepage
\section{Videos}
\begin{figure}[H]
\centering
\subfloat[\label{subvid:trat-l}]{\begin{minipage}[t]{\linewidth}
  \centering
  \includegraphics[width=.8\linewidth]{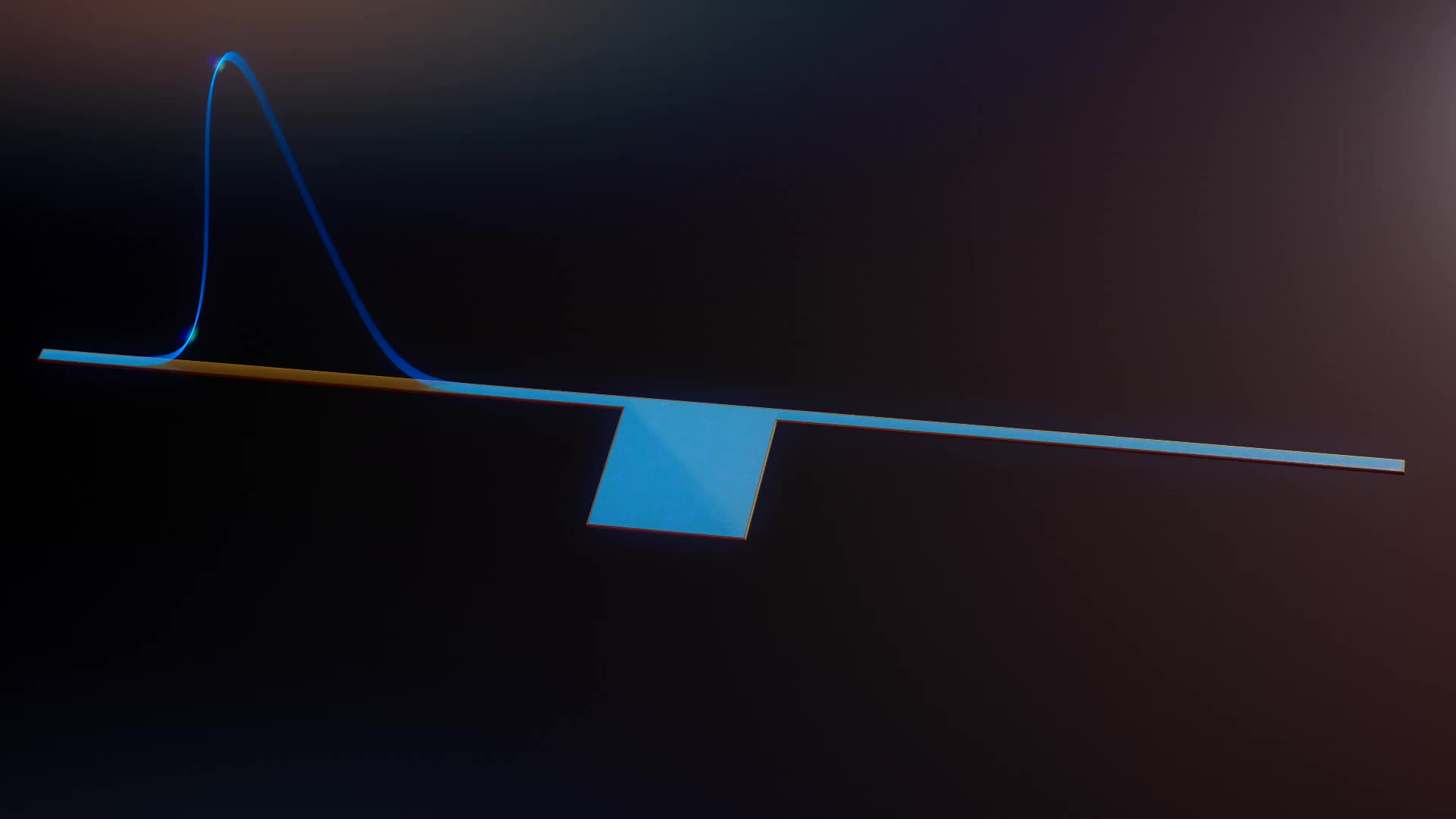}
\end{minipage}}\\
\subfloat[\label{subvid:trat-r}]{\begin{minipage}[t]{\linewidth}
  \centering
  \includegraphics[width=.8\linewidth]{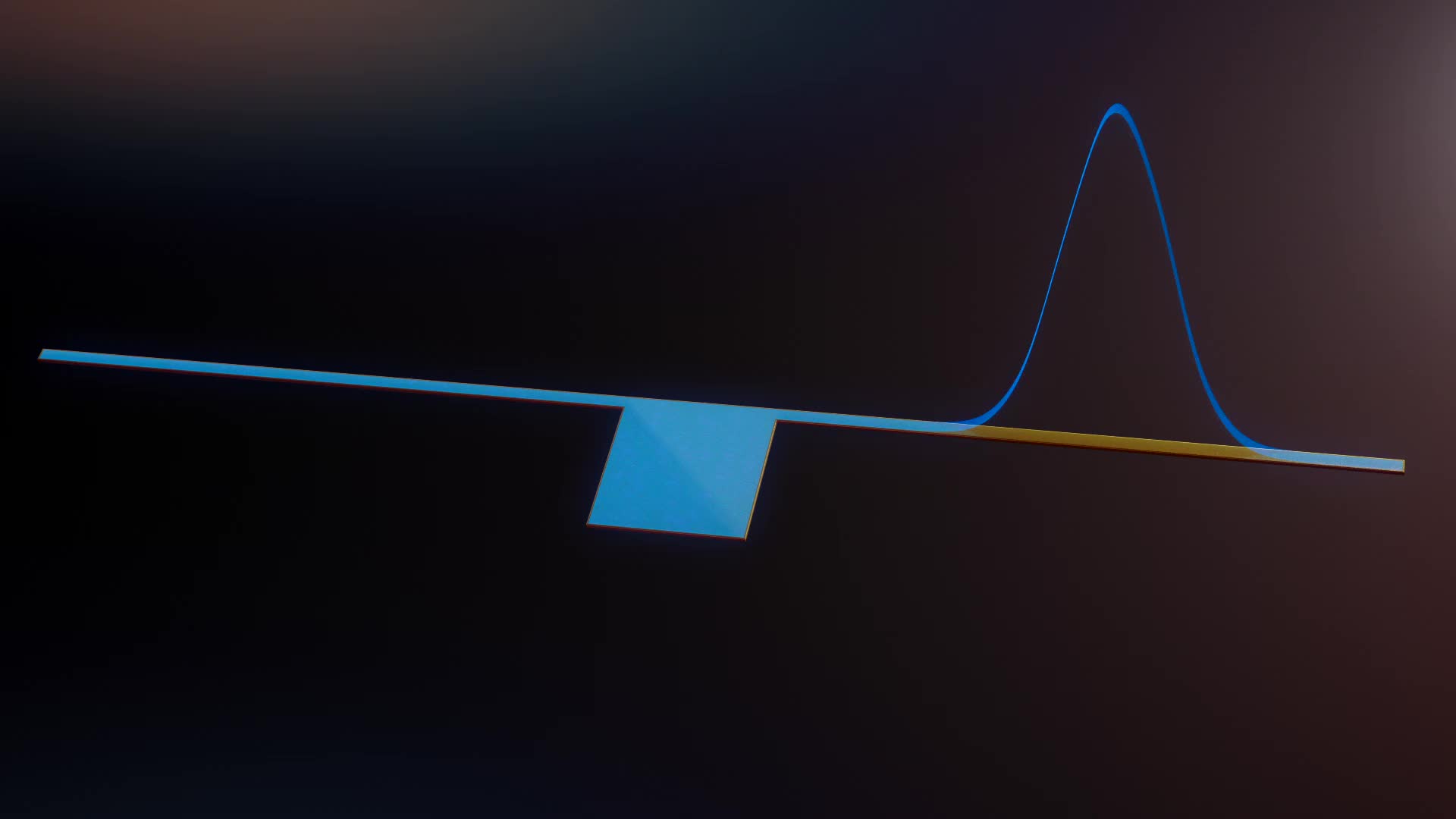}
\end{minipage}}
\caption{\subref{subvid:trat-l} Unitary propagation of a wave packet (blue) across a conducting line (gold) with a transverse asymmetry. A magnetic field is applied perpendicular to the conducting plane. The wave packet, presented as $|\psi(r,t)|^2$, arrives from the left and is partially reflected. The data have been obtained by exact diagonalization as described in the main text. \subref{subvid:trat-r} The wave packet arriving from the right.}
\label{vid:trat}
\end{figure}

\begin{figure}[H]
\centering
\subfloat[\label{subvid:amzi-l}]{\begin{minipage}[t]{\linewidth}
  \centering
  \includegraphics[width=.8\linewidth]{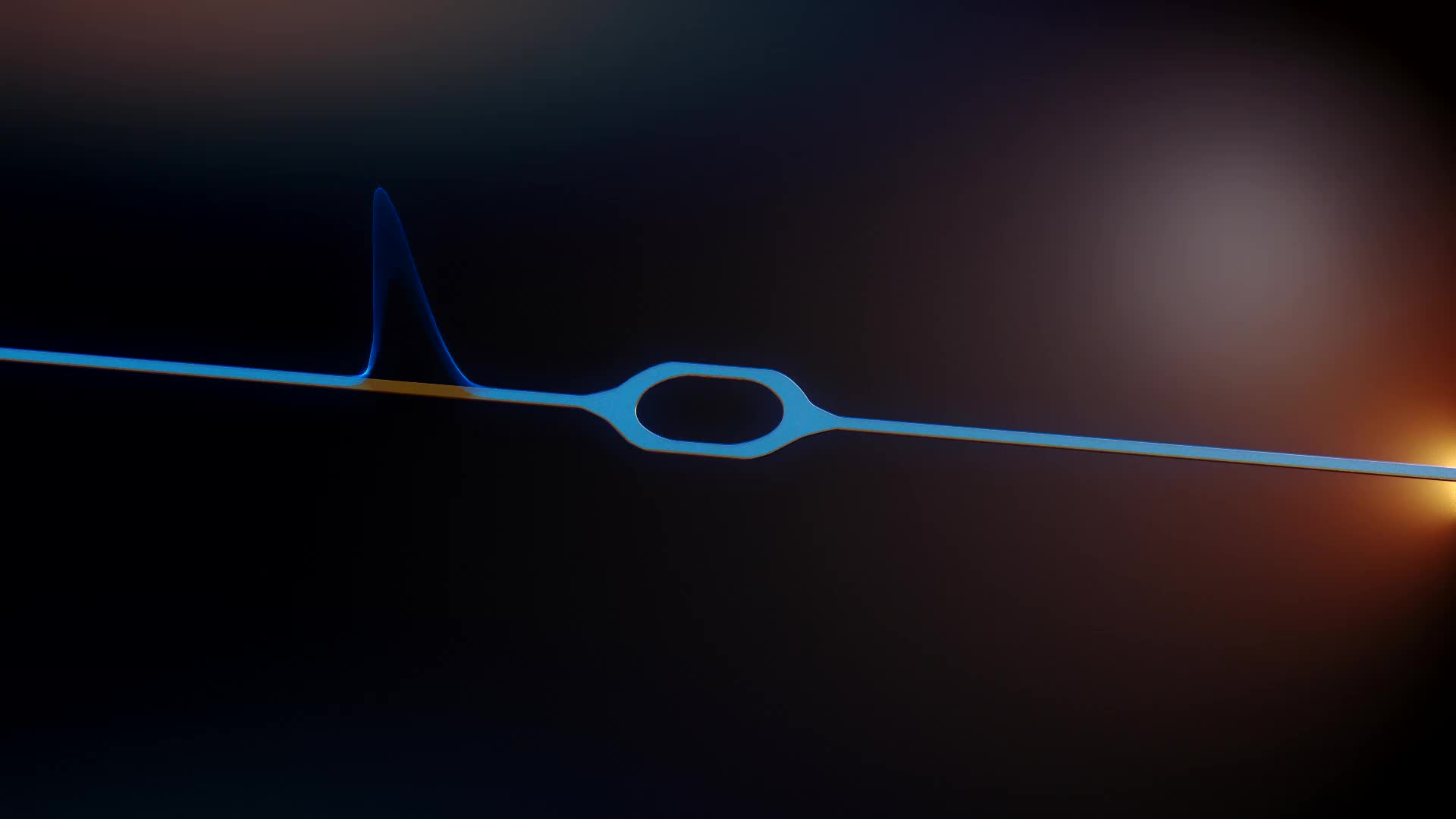}
\end{minipage}}\\
\subfloat[\label{subvid:amzi-r}]{\begin{minipage}[t]{\linewidth}
  \centering
  \includegraphics[width=.8\linewidth]{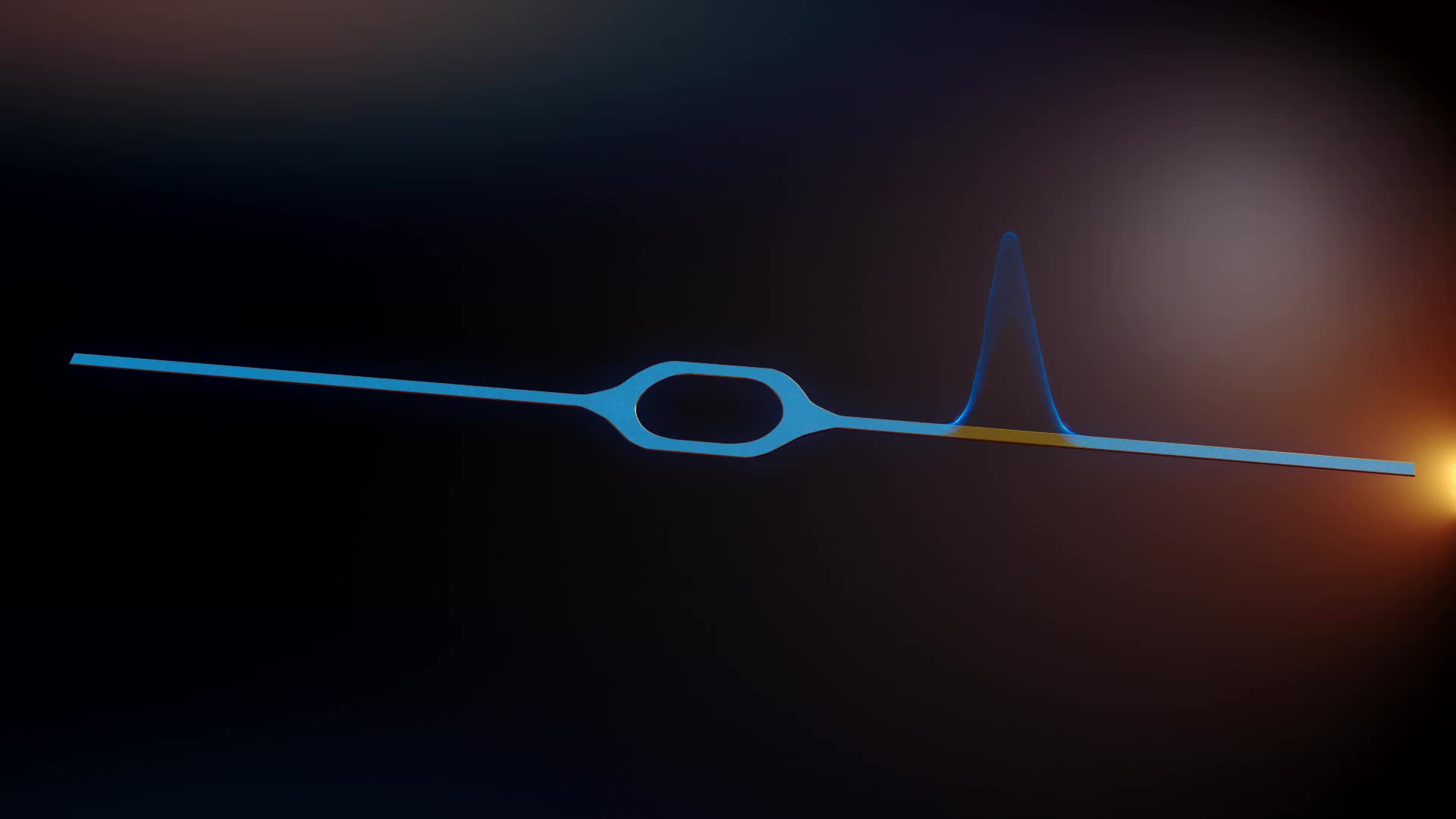}
\end{minipage}}
\caption{\subref{subvid:amzi-l} Unitary propagation of a wave packet (blue) across an asymmetric Aharonov-Bohm ring (gold) biased with a magnetic flux penetrating the hole of the ring. The wave packet, presented as $|\psi(r,t)|^2$, arrives from the left and passes the ring in a straightforward manner. The data have been obtained by exact diagonalization as described in the main text. \subref{subvid:amzi-r} The wave packet arriving from the right passes the ring only after having been reflected back and forth.}
\label{vid:amzi}
\end{figure}

\begin{figure}[H]
\centering
\includegraphics[width=.8\linewidth]{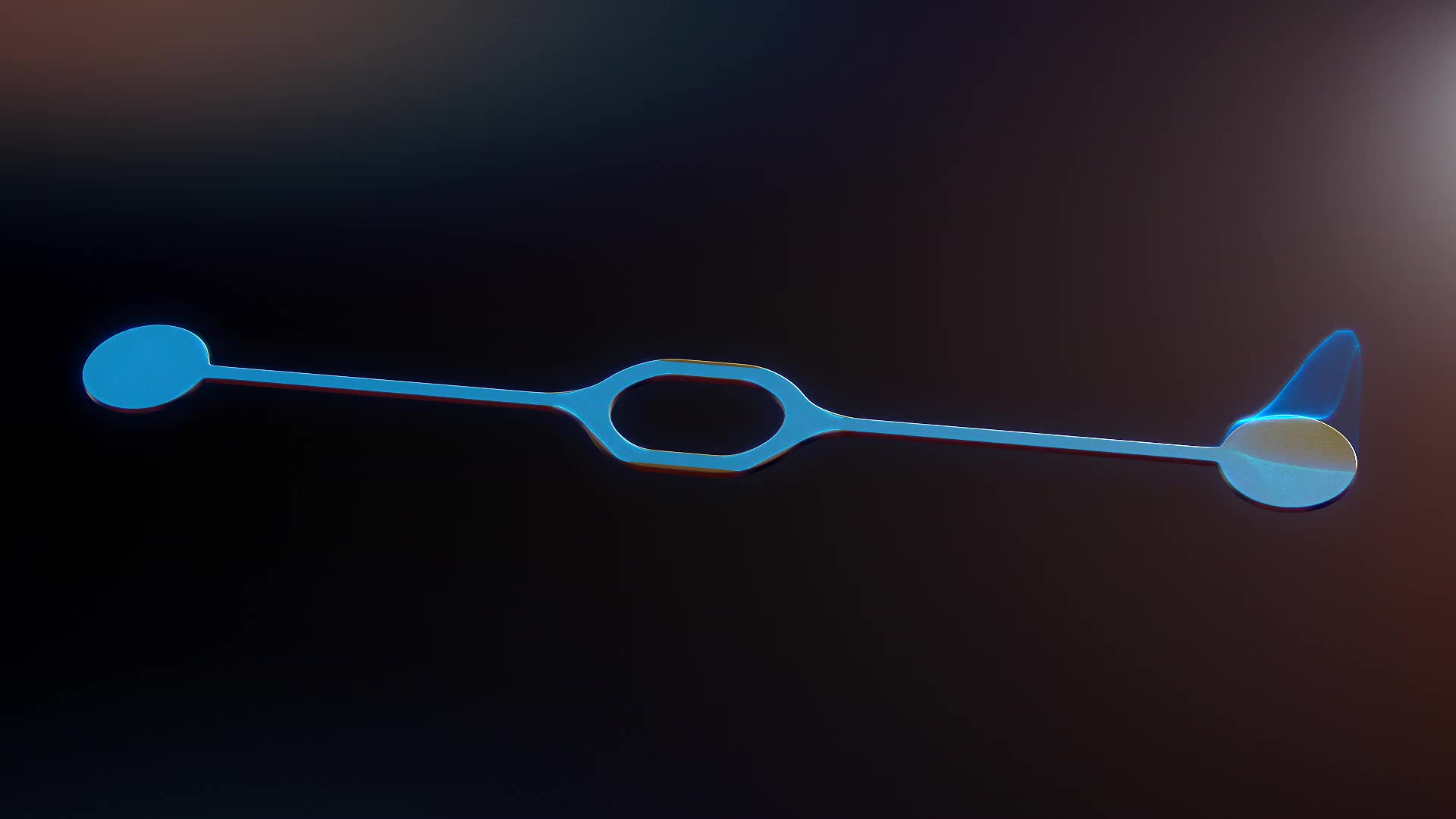}
\caption{Propagation of a wave packet in a closed system provided by two contacts and an asymmetric Aharonov-Bohm ring biased with a magnetic flux penetrating the hole of the ring. The wave packet is represented by $|\psi(r,t)|^2$. The unitary propagation is interrupted by three collapse processes highlighted in purple which correspond to negative and to positive result measurements. The data have been obtained by the method described in the main text.}
\label{vid:amzi-closed}
\end{figure}

Videos of selected time evolutions.
Fig.~\ref{vid:trat} shows the starting configuration of the time evolution of a wave packet passing a device unitarily with a transverse asymmetry that is shown in the videos.
Fig.~\ref{vid:amzi} shows the corresponding configuration for an asymmetric Aharonov-Bohm ring.
Fig.~\ref{vid:amzi-closed} shows the starting image of the video display the wave packet propagating across an asymmetric Aharonov-Bohm ring interrupted by collapse processes.
The videos are available at \url{https://fkf.mpg.de/mannhart}. 
\vfill

\vfill

\cleardoublepage
\section{Time-dependent wave packets and currents}
\label{sec:currents}

In this part of the supplement we show that the time-dependent currents of wave packets are not reciprocal in a two-terminal device. Nevertheless, the total transported charge is reciprocal in the long time limit. Unitary evolution yields therefore reciprocal transport probabilities independent from the shape of the electronic wave function in the steady state.

We consider a one-dimensional wire along the $x$-axis on a two-dimensional substrate with a local potential $V(\mathbf{\hat{r}})$ confined to a finite region in the $x$/$y$-plane and a homogeneous magnetic field perpendicular to the plane. The wire is attached to charge reservoirs (ports) on the left and the right side with chemical potentials $\mu_l$ and $\mu_r$, respectively. The single-particle Hamiltonian reads then (with electron charge $q_e=e$ and mass $m_e$)
\begin{align}
H = \frac{\hat{p} - \frac{e}{c}\mathbf{A}(\mathbf{\hat{r}})^2}{2m_e} + V(\mathbf{\hat{r}}) + V_\mathrm{conf}(\mathbf{\hat{r}}).
\label{eqn:schroedinger}
\end{align}
The potential ${V_\mathrm{conf}(\mathbf{\hat{r}})}$ confines the electrons to the wire, whereas the potential ${V(x,y)=0}$ for ${|x|>R}$. With the choice ${\mathbf{A}(x,y)=(0,Bx,0)^T}$, we can approximate the eigenfunctions of \eqnref{schroedinger} in the wire outside the interaction region by plane waves ${\psi^{1,2}(x,y)=\phi_n(y)\psi_k^{1,2}(x)}$
with $\psi_k^{1,2}(x)\propto\exp(\pm ikx)$ The transversal quantum number is denoted by $n$. We assume for simplicity that the energy ${E_k = \hbar^2k^2/(2m_e)}$ of ${\psi_{k,n}^{1,2}}$ does not depend on it.

The asymptotic form of ${\psi_k^{1,2}(x)}$ for ${|x|\gg R}$ reads in general
\begin{align}
\psi_k^1 &= \frac{1}{N} \begin{cases}
e^{ikx} + R_k^l e^{-ikx} & x \ll -R \\
T_k^l e^{ikx} & x \gg R
\end{cases}
\label{eqn:asympt-psi1}
\\
\psi_k^2 &= \frac{1}{N} \begin{cases}
T_k^r e^{-ikx} & x \ll -R \\
e^{-ikx} + R_k^r e^{ikx} & x \gg R
\end{cases}
\label{eqn:asympt-psi2}
\end{align}
with the normalization factor ${N\propto\sqrt{L}}$, where ${L\gg R}$ is the length of the wire. A wave function with energy $E(k)$ is in general a superposition of incoming waves from the left with amplitudes $A_k^-$ and from the right with amplitudes ${A_k^+(k\geq 0)}$. Both waves are scattered in the region of ${V(r)\neq 0}$, with direction-dependent transmission and reflection amplitudes $T_k^s$, $R_k^s$, ${s=r,l}$. The S-matrix of the system reads
\begin{align}
\hat{S}_k = \begin{pmatrix}
R_k^l & T_k^r \\ T_k^l & R_k^r
\end{pmatrix}.
\end{align}
Unitarity of $\hat{S}_k$ requires:
\begin{align}
\big|T_k^s\big|^2 &= 1-\big|R_k^s\big|^2 \quad s=r,l\\
\big|T_k^l\big|^2 &= \big|T_k^r\big|^2.
\label{eqn:reciprocity}
\end{align}
In the steady state, the current $j_{k,n}(x_0)$ of state $\psi^1_{k,n}$ injected from the left reservoir at some point $x_0\ll -R$ reads
\begin{align}
j_{k,n}^l(x_0) = \frac{ev(k)}{N^2},
\end{align}
with the electron velocity $v(k)=\hbar k/m_e$. The total current injected from the left and passing the point $x_0$ is
\begin{align}
I^l(x_0) = e\sum\limits_{k,n}v(k)N^{-2} = eN_y\int\limits_0^{\mu_l}\diff E \rho\left(E\right)v\left(k\left(E\right)\right)L^{-1}.
\end{align}
Here, $N_y$ is the number of transversal channels and $\rho(E) = 2L/\left(hv(E)\right)$ is the one-dimensional density of states per channel. It follows that
\begin{align}
I^l(x_0) = \frac{2e}{h}N_y\mu_l.
\end{align}
The total current at $x_0$ contains a contribution from electrons which are reflected at the central region and those transmitted from the right reservoir:
\begin{align}
I(x_0) = \frac{2e}{h}N_y\left(
\int\limits_0^{\mu_l}\diff E \left|T_{k(E)}^l\right|^2 - \int\limits_0^{\mu_r}\diff E \left|T_{k(E)}^r\right|^2\right).
\end{align}
To obtain the linear response for stationary states, one averages ${\left|T_{k(E)}^{l,r}\right|^2}$ over $E$. Using \eqnref{reciprocity} we obtain \cite{phas-coher-trans},
\begin{align}
I(x_0)
= \frac{2e}{h}N_y\overline{\left|T_{k(E)}^l\right|^2}(\mu_l-\mu_r)\\
= \frac{2e^2}{h}N_y\overline{\left|T_{k(E)}^l\right|^2}(V_l-V_r),
\label{eqn:landauer-argument}
\end{align}
where $V_l$ ($V_r$) are the voltages of the left (right) reservoir. One sees that because of \eqnref{reciprocity}, the coherent linear response for stationary states is reciprocal and $I(x_0) = 0$ if $V_l = V_r$.

Now we consider wave packets instead of the time-independent stationary eigenstates of the system, e.g., of Gaussian form. At the initial time $t=0$, the packet is localized in the left part of the wire around $x^l_\mathrm{in} \ll -R$ with momentum expectation value $\hbar k_0>0$, moving to the right.
\begin{align}
\psi_l(x,0)&=(\pi \sx^2)^{-1/4}e^{-\frac{(x-x_\mathrm{in}^l)^2}{2\sx^2}+ik_0x}\\
&= \int\limits_0^\infty\diff k\left(A^-_{l,k}\psi^1_k(x)+A^+_{l,k}\psi^2_k(x)\right).
\label{eqn:packet-psi}
\end{align}
If the packet is sufficiently narrow, the coefficients $A^{\pm}_{l,k}$ may be computed using the asymptotic expressions given in \eqnref{asympt-psi1} and \eqnref{asympt-psi2} to obtain
\begin{align}
A^-_{l,k} &= \frac{\sx^{1/2}N}{\sqrt{2}\pi^{3/4}}e^{-\frac{\sx^2}{2}(k-k_0)^2-ikx_\mathrm{in}^l} \\
A^+_{l,k} &= (B^-_{l,k}-A^-_{l,k}R^l_k)/T^r_k\\
&B^-_{l,k} = \frac{\sx^{1/2}N}{\sqrt{2}\pi^{3/4}}e^{ik_0x^l_\mathrm{in}}e^{-\frac{\sx^2}{2}(k+k_0)^2+ikx^l_\mathrm{in}}.
\end{align}
The state given in \eqnref{packet-psi} is time-dependent. The associated current at some point $x_r\gg R$ on the right side of the interacting region reads at time $t$
\begin{align}
j_l(x_r,t) = \frac{e}{m_e}\mathbf{\Re}\big\langle\psi_l(t)\big|\delta(x-x_r)\hat{p}_x\big|\psi_l(t)\big\rangle,
\label{eqn:current-def}
\end{align}
where $\mathbf{\Re}z$ denotes the real part of $z$. Now
\begin{align}
\psi_l(x,t) = \int\limits_0^\infty\diff k\left(A^-_{l,k}\psi^1_k(x) + A^+_{l,k}\psi^2_k(x)\right)e^{-iE(k)t/\hbar},
\end{align}
and to compute $j_l(x_r,t)$ we may use the asymptotics of $\psi_k^{1,2}(x)$. This yields
\begin{align}
j_l(x_r,&t)=\frac{e\hbar}{m_eN^2}\mathbf{\Re}
\int\limits_0^\infty\diff k
\int\limits_0^\infty\diff k'\bigg\{k'\big[
C^*_{l,k}C_{l,k}e^{i(k'-k)x_r} \nonumber\\
&+A^{+*}_{l,k}C_{l,k}e^{i(k+k')x_r}\big]
-k'\big[
C^*_{l,k}A^+_{l,k}e^{-i(k+k')x_r}\nonumber\\
&+ A^{+*}_{l,k}A^+_{l,k}e^{i(k-k')x_r}\big]
\bigg\}
e^{i\left(E(k)-E(k')\right)t/\hbar},
\label{eqn:pack-current-l}
\end{align}
with ${C_{l,k}=T^l_k A^-_{l,k} + A^+_{l,k}R^r_k}$. We consider now a second initial state $\psi_r(x,0)$ obtained from $\psi_l(x,0)$ by reflection of $x$ at the origin: ${\psi_r(x,0)=\psi_l(-x,0)}$, centered around ${-x^l_\mathrm{in}\gg R}$.
\begin{align}
\psi_r(x,0)&=(\pi \sx^2)^{-1/4}e^{-\frac{(x+x^l_\mathrm{in})^2}{2\sx^2}-ik_0 x}\\
&=\int\limits_0^\infty\diff k\left(A^-_{r,k}\psi^1_k(x)+A^+_{r,k}\psi^2_k(x)\right).
\end{align}
The state $\psi_r(x,t)$ has average momentum $-\hbar k_0$ and moves to the left. The coefficients $A^{\pm}_{r,k}$ read
\begin{align}
A^+_{r,k} &= A^-_{l,k},\\
A^-_{r,k} &= (B^+_{r,k}-A^+_{r,k}R^r_k)/T^l_k,\quad
B^+_{r,k} = B^-_{l,k}.
\end{align}
For this state, we calculate the current at a time $t$ and at the point $x_l=-x_r$. The result is
\begin{align}
j_r(-x_r,&t)=\frac{e\hbar}{m_e N^2}\mathbf{\Re}
\int\limits_0^\infty\diff k
\int\limits_0^\infty\diff k'\bigg\{k'\big[
A^{-*}_{r,k}A^-_{r,k}e^{i(k-k')x_r} \nonumber\\
&+C^*_{r,k}A^-_{r,k}e^{-i(k+k')x_r}\big]
-k'\big[
A^{-*}_{r,k}C_{r,k}e^{i(k+k')x_r} \nonumber\\
&+ C^*_{r,k}C_{r,k}e^{i(k'-k)x_r}\big]
\bigg\}e^{i\left(E(k)-E(k')\right)t/\hbar},
\label{eqn:pack-current-r}
\end{align}
with ${C_{r,k}=T^r_kA^+_{r,k}+A^-_{r,k}R^l_k}$. By comparing \eqnref{pack-current-l} and \eqnref{pack-current-r}, one sees that the currents are reflection-symmetric, i.e., ${j_r(-x_r,t)=-j_l(x_r,t)}$, if ${T^l_k=T^r_k}$ and ${R^l_k=R^r_k}$. This follows from the reflection symmetry of the Hamiltonian
\begin{align}
&H(\hat{p}_x,\hat{p}_y,\hat{x},\hat{y},
  \mathbf{A}(\hat{x},\hat{y}))\nonumber\\
&\quad= \mathscr{R}(H)\nonumber\\
&\quad= H(-\hat{p}_x,\hat{p}_y,-\hat{x},\hat{y},\mathbf{A}(-\hat{x},\hat{y})),
\label{eqn:mirror-symmetry}
\end{align}
because then
\begin{align}
\mathscr{R}(\psi)(t)
&=e^{itH/\hbar}\mathscr{R}(\psi(0))\nonumber\\
&=\mathscr{R}\left(e^{itH/\hbar}\psi(0)\right)
=\mathscr{R}\big(\psi(t)\big).
\end{align}
However, if \eqnref{mirror-symmetry} is not satisfied, we have in general
\begin{align}
T^l_k = T^r_k e^{i\theta_k},\quad R^l_k=R^r_k e^{i\vartheta_k}.
\end{align}
The left and right transmission and reflection coefficients differ by phase factors, which are allowed by the unitarity of the S-matrix $\hat{S}_k$. If the $\theta_k$, $\vartheta_k$ do not vanish, it follows ${j_r(-x_r,t)\neq-j_l(x_r,t)}$, i.e., the time-dependent currents are not reciprocal.

Nevertheless, the total charge transported from the left to the right over a sufficiently long time equals the total charge flowing from the right to the left, so that the steady state has reciprocal transport characteristics, in accordance with the result \eqnref{landauer-argument}, which follows from \eqnref{reciprocity}. The charge of initial state $\psi_l(x,0)$ flowing through the point $x_r$ in the time interval $[0,T]$ to the right is given by
\begin{align}
Q_{l\rightarrow r}(x_r,T)=\int\limits_0^T\diff t j_l(x_r,t),
\label{eqn:charge-integral-lr}
\end{align}
whereas the charge of state $\psi_r(x,0)$ flowing to the left through point $-x_r$ reads
\begin{align}
Q_{r\rightarrow l}(-x_r,T)=-\int\limits_0^T\diff t j_r(-x_r,t).
\label{eqn:charge-integral-rl}
\end{align}
The limit ${T\rightarrow\infty}$ exists ${(0\leq Q(x,T)\leq e)}$ because we consider an open system without periodic boundary conditions. Thus the electron may pass the point $x$ and never come back - otherwise $Q(x,\infty)$ would be either zero or infinity. Reciprocity of the steady state corresponds to ${Q_{l\rightarrow r}(x_r,\infty)=Q_{r\rightarrow l}(-x_r,\infty)}$. Because
\begin{align}
\lim\limits_{T\rightarrow\infty}\int\limits_0^T\diff te^{itE/\hbar}=\pi\delta(E/\hbar)+i\mathscr{P}\left(\frac{\hbar}{E}\right),
\end{align}
where $\mathscr{P}$ denotes the principal part, we obtain
\begin{align}
Q_{l\rightarrow r}(x_r,\infty)=\frac{e\pi}{N^2}\int\limits_0^\infty\diff k\left[\big|C_{l,k}\big|^2-\big|A^+_{l,k}\big|^2\right],
\label{eqn:charge-lr}
\end{align}
and
\begin{align}
Q_{r\rightarrow l}(-x_r,\infty)=\frac{e\pi}{N^2}\int\limits_0^\infty\diff k\left[\big|C_{r,k}\big|^2-\big|A^-_{r,k}\big|^2\right].
\label{eqn:charge-rl}
\end{align}
Using the identity ${T^{r*}_k R^l_k+T^l_k R^{r*}_k=0}$, which follows from the unitarity of $\hat{S}_k$, one may show that the integrands of \eqnref{charge-lr} and \eqnref{charge-rl} are the same. Therefore
\begin{align}
Q_{l\rightarrow r}(x_r,\infty)=Q_{r\rightarrow l}(-x,\infty),
\end{align}
as anticipated.

\section{Numerical implementation}
\label{sec:numerics}
\subsection{Lindblad formalism}

The numerical implementation starts with the tight-binding Hamiltonian:
\begin{align}
H_0=4t\left(\sum\limits_{i\in\Lambda}c_i^\dagger c_i\right)-t\left(\sum\limits_{\langle i,j\rangle\in\Lambda} c_i^\dagger c_j+\text{h.c.}\right),
\end{align}
\begin{figure}
  \centering
  \includegraphics[scale=1]{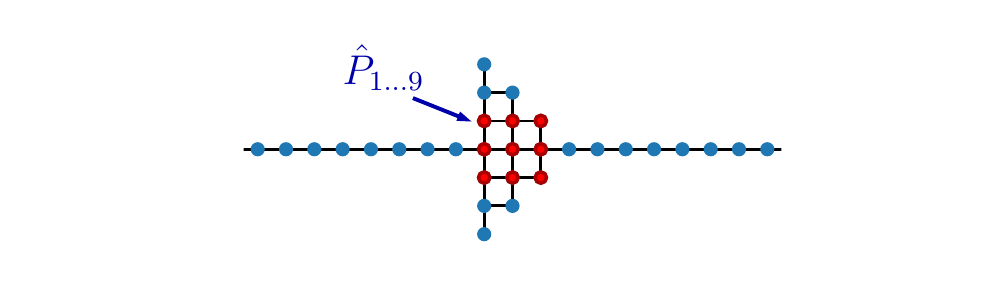}
  \caption{Sketch of the tight-binding implementation of the device A. The circles represent the system sites, the red sites correspond to inelastic scattering centers. These centers mediate the coupling to the environment.}
  \label{fig:scatter-sites}
\end{figure}
where $t$ denotes the hopping energy, $\langle i, j\rangle$ a pair of neighboring sites and $c^\dagger_i$, $c_i$ the creation and annihilation operators on site $i$. In the following, we confine ourselves  to the single-particle subspace. For $t$ we take \SI{1}{\eV}, corresponding to typical dwelling times of several \si{\fs} in the device. The lattice $\Lambda$ defines the system geometry and size. The following paragraphs refer to the systems used in Sections \ref{sec-dyn} and \ref{sec-decoherence}.
The implementation details of the minimized system shown in Section \ref{sec-steady} can be found in Section \ref{sec:minisys} of this appendix.
The leads are implemented as long lines. The systems contain up to 5065 sites to ensure that during the observation time $t_{\text{fin}}$ the wave packets do not return back into to device. Gaussian wave packets with ${\sx=3a}$ and ${8a}$ and $k_0=\pi/(3a)$, where $a=\SI{0.3}{nm}$ is the lattice constant of $\Lambda$, serve as initial wave functions for the devices A, B, D and C, respectively. We start from the density matrix of a pure state
$\rho_e(0)=|\phi\rangle\langle\phi|$ where $|\phi\rangle$ is either a right-moving wave packet placed in the left port $L$, $|\phi^{L}_{k_0}\rangle$, or a left-moving packet, $|\phi^{R}_{-k_0}\rangle$ located in the right port $R$. $\rho_e(t)$ is then numerically evolved with the Lindblad equation \eqnref{lindblad1} to compute the quantities
$\PLR$ and $\PRL$ as function of $\Gamma$. For the devices A, B, and D we have used nine different scattering centers located at the sites indicated in \figref{scatter-sites} with the same coupling $\gamma_{\bm{r}}=\Gamma$. For the device C we regard the entire lower interferometer arm as a scattering center. In this case, there is only one projection operator onto the subspace spanned by all sites in the lower interferometer arm. The layout of the devices is further illustrated in Fig.~\ref{fig:tbsys}.

\subsection{Monte-Carlo wave functions}
As in \eqnref{current-def}, we calculate the time-dependent probability currents at sites $\bm{r}_L=(-25,0)$ and $\bm{r}_R=(25,0)$ to obtain the total transmitted and reflected charges as in \eqnref{charge-integral-lr} and \eqnref{charge-integral-rl}.
The time is discretized in steps of width ${\Delta t=\min\{\SI{2.6e-17}{s}, \tau_c/20\}}$ to obtain an adequate temporal resolution in the time integrals of the currents and the dynamical collapse events.

The collapse events are computed in the following way: A collapse occurs with probability $p_c$ per unit time, the rate constant of the corresponding Poisson process is then $1/\tau_c=p_c$, where $\tau_c$ is the average time between collapses occurring at times $t_{c,j}$ and $t_{c,j+1}$. The rate constant $1/\tau_c$ is proportional to the coupling between the electron and the localized degree of freedom with which it interacts inelastically. The inelastic event itself is treated like a measurement process: The electron with wave function $\psi(t_{c,j})$ becomes localized (positive measurement result) and acquires one of the wave functions $\psiloci$ at time $t_{c,j}+\epsilon$ with probability $p_i=|\langle\psi(t_{c,j})|\psiloci \rangle|^2$. The index $i$ runs over the inelastic scattering centers as depicted in \figref{scatter-sites}.

In case of a negative measurement result, the state at $t_{c,j}+\epsilon$ is the projection of $\psi(t_{c,j})$ onto the orthogonal complement $\psilocperp$ of the sum of all $\psiloci$. This happens with probability $1-\sum_i p_i$. From the time $t_{c,j}+\epsilon$ onwards, the state develops according to the time-dependent Schrödinger equation until the next collapse event at time $t_{c,j+1}$, where the wave function changes again discontinuously.

For Figs. \ref{fig:colormaps}, \ref{fig:colormaps-sdot} we use the following collapse scenario: $\psiloci=|\bm{r}\rangle$, $\bm{r}$ being one of the nine lattice points
$\{(0,0), (\pm 1,0), (0, \pm 1), (\pm 1, \pm 1), (\pm 1, \mp 1)\}$. In case of a negative measurement, the wave function is projected onto the orthogonal complement of the span of all nine $|\bm{r}\rangle$.

We demonstrate now that averaging over the stochastic trajectories $|\psi_l(t)\rangle$ obtained in this way yields the same result for the reduced density matrix $\rho_e(t)$ as computed via the Lindblad equation \eqnref{lindblad1}.
First, we consider the case of a single scattering center at site $\bm{r}$. Let's assume that for the $l$-th trajectory the wave function at time $t$ is $|\psi_l(t)\rangle$. At time $t+\delta t$, the wave function reads then
\begin{widetext}
\begin{equation}
  |\psi_l(t+\delta t)\rangle=\left\{
  \begin{array}{ccl}
    \left(1-i\frac{\delta t}{\hbar}H_0\right)|\psi_l(t)\rangle & \text{with probability} & 1-\delta t p_c,\\
    |\bm{r}\rangle    & '' & \delta t p_c|\langle \psi_l(t)|\bm{r}\rangle|^2,\\
    |\bm{r}^\perp\rangle & '' & \delta t p_c(1-|\langle \psi_l(t)|\bm{r}\rangle|^2).
  \end{array}
  \right.
  \label{eqn:probab}
\end{equation}
\end{widetext}
This entails the following mixed density matrix $\rho_l$ at time $t+\delta t$ belonging
to the $l$-th trajectory up to time $t$ (neglecting terms of order $(\delta t)^2$),
\begin{align}
  \rho_l(t+\delta t) =&
  (1-\delta t p_c)|\psi_l(t)\rangle\langle\psi_l(t)| \nonumber\\
  &-\delta t\frac{i}{\hbar}[H_0,|\psi_l(t)\rangle\langle\psi_l(t)|]\nonumber\\
  &+ \delta t p_c \hat{P}_{\bm{r}}|\psi_l(t)\rangle\langle\psi_l(t)|\hat{P}_{\bm{r}} \nonumber\\
  &+ \delta t p_c(\ident - \hat{P}_{\bm{r}})|\psi_l(t)\rangle\langle\psi_l(t)|(\ident - \hat{P}_{\bm{r}}),
  \end{align}
which yields the following equation for the derivative of $\rho_l(t)$ at time $t$,
\begin{align}
\left.\frac{\diff \rho_l(t')}{\diff t'}\right|_t
= &-\frac{i}{\hbar}[H_0,\rho_l(t)] \nonumber\\
  &+ p_c\left(2\hat{P}_{\bm{r}}\rho_l(t)\hat{P}_{\bm{r}} - \hat{P}_{\bm{r}}
\rho_l(t) - \rho_l(t)\hat{P}_{\bm{r}}\right).
\end{align}
The average over all trajectories $\rho(t)={\cal N}^{-1}\sum_l\rho_l(t)$ therefore satisfies the same differential equation for all times, from which we obtain the Lindblad equation \eqnref{lindblad1} for a single scattering center with the identification $p_c=1/\tau_c=\gamma_{\bm{r}}/2$. The generalization to several scattering centers is straightforward.

The presented stochastic unraveling of the Lindblad equation is not unique. Another stochastic process, equivalent to \eqnref{lindblad1}, considers only positive measurements, the ``quantum jump'' projects the wave function with probability $p_c$ always onto one of the $\psiloci$, but never onto $\psilocperp$. To account for null measurements, the deterministic evolution of $|\psi_l(t)\rangle$ between collapse events proceeds not with the hermitean Hamiltonian $H_0$ but with the non-hermitean operator $H=H_0-ip_c\hat{P}_{\bm{r}}$. For this, an additional normalization of $|\psi_l(t)\rangle$ during the evolution is required because $H$ does not conserve the norm of the wave function \cite{carmichael2009open}. The corresponding master equation for $\rho(t)$ reads then
\begin{align}
\frac{\diff \rho}{\diff t}= -\frac{i}{\hbar}(H\rho -\rho H^\dagger) + 2p_c\hat{P}_{\bm{r}}\rho\hat{P}_{\bm{r}},
\end{align}
which is again \eqnref{lindblad1} \cite{molmer1993}.

The accuracy of the calculations of the unitary dynamics in our numerical implementation is limited by the finite size of the leads, which sets an upper bound to the observation time $t_{\text{fin}}$ due to recurrence of the waves after they have been reflected by the lead ends. In our case, $t_{\text{fin}}=\SI{1.58}{ps}$.

The statistical accuracy of the collapse dynamics obviously increases with the number of sampled trajectories. In the calculations, a minute fraction of trajectories had to be discarded because accumulation of numerical discretization errors led to division-by-zero errors or minute negative probabilities. For \figref{colormaps}, \lratValidTrajectories{} trajectories were used for each transmission map, \lratDiscardedTrajectories{} trajectories were discarded. The same discretization errors cause the calculated total probability of all trajectories to deviate from 1. On average, the probability conservation violation equals \num{5e-4}. For \SI{99}{\percent} of the trajectories it is better than \num{1e-2}, the largest violation being \num{0.1}. A third systematic error concerns the fact that at $t_0$ the wave function is not completely zero inside the device, such that $Q_A(t_0) + Q_B(t_0) \neq 1$, where $Q_A$, $Q_B$ are the charges in the contacts. Less than \num{1e-2} of the trajectories leave a residual charge $Q_A(t_0)+Q_B(t_0)-1 > \num{1e-3}$ (all numbers referring to the calculations shown in \figref{colormaps}).

Figure~\ref{fig:colormaps-sdot} shows, analogous to Fig.~\ref{fig:colormaps}, the transmission probability histograms of the symmetric device shown in Figs.~\ref{subfig:sys-sdot},~\ref{subfig:tbsys-sdot}. No sorting is observed.

\subsection{Minimized model used to calculate steady states}
\label{sec:minisys}

The calculations of steady states in Section \ref{sec-steady} were done using the smaller tight-binding system sketched in \figref{lindblad-chain-sys}. The Hamiltonian of this system is given by
\begin{align}
H_0=\left(\sum\limits_{i\in\Lambda}\left(2t+V(i)\right)c_i^\dagger c_i\right)
   - t\left(\sum\limits_{\langle i,j\rangle\in\Lambda} c_i^\dagger c_j + \text{h.c.}\right),
\end{align}
where the tight-binding lattice $\Lambda$ is given by ${\Lambda=\{\ket{1},\ket{2}\dots\ket{9}\}}$, ${\langle i,j \rangle}$ are again the pairs of neighboring sites. Sites that are connected by solid lines in \figref{lindblad-chain-sys} are of course neighbors. In the ring configuration used to demonstrate the existence of persistent currents, the sites at the left and right ends of the chain are nearest neighbors as well. The $c_i^\dagger$ and $c_i$ are the creation and annihilation operators on site $i$ and $V(i)$ is the electric potential at site $i$. The asymmetric potential barrier used for the calculations shown in the main text is given by
\begin{align}
  V_\mathrm{asym}(i) = \setstretch{1.1}\begin{cases}
    ~3t &\text{if } i=4\\
    ~2t &\text{if } i=5\\
    ~t &\text{if } i=6\\
    ~0 &\text{else}
  \end{cases}.
  \label{eqn:Vasym}
\end{align}
The symmetric potential barrier used in \figref{lindblad-thermo-sdot} is given by
\begin{align}
  V_\mathrm{sym}(i) = \setstretch{1.1}\begin{cases}
    ~2t &\text{if } 4\leq i\leq6\\
    ~0 &\text{else}
  \end{cases}.
\end{align}
The two wave packet states used for the jump operators are given by
\begin{align}
  \ket{\psi_{\bm{r}}({\pm\bm{p}})} = \frac{1}{\sqrt{6}}\left(0,0,0,
        e^{\pm i\frac{\pi}{3}},2,e^{\mp i\frac{\pi}{3}},
        0,0,0\right)^\mathrm{T}
\end{align}
The electrical current $J$ associated with a density operator $\rho$ is calculated using the velocity operator $\hat{v}$,
\begin{align}
  J &= -en\tr(\hat{v}\rho),\\
  \hat{v} &= -\frac{i}{\hbar}[\hat{x},H]
\end{align}
with the elementary charge $e$ and the 1D carrier density $n$ and the position operator $\hat{x}$. The density is chosen such that the chain contains one carrier $n=1/(9a)=\SI{3.7e6}{\per\cm}$ with the lattice constant ${a=\SI{.3}{\nm}}$.

The thermal relaxation processes mentioned in the last paragraph of Section \ref{sec-steady} are implemented by additional jump operators ${\hat{B}_{ij}=|E_j\rangle\langle E_i|}$, where \ket{E_i} is the eigenstate of the Hamiltonian $H_0$ with energy $E_i$. The corresponding rates are given by ${\gamma_{ij}=\gamma_\mathrm{th}\exp(-E_j/kT)}$. The resulting Lindblad equation reads
\begin{equation}
  \frac{\diff \rho_e}{\diff t} = -\frac{i}{\hbar}[H_0,\rho_e]
  +\sum_{\bm{r}}\gamma_{\bm{r}}\cD[\hat{A}_{\bm{r}}](\rho_e) +
  \sum_{i,j}\gamma_{ij}\cD[\hat{B}_{ij}](\rho_e).
  \label{eqn:lindblad3}
\end{equation}
\figref{lindblad-thermal-jumps} shows the temporal behavior of the charge imbalance
for the chain with the thermalizing jump operators.
The initial state $\rho_e(0)$ is the unique steady state of \eqnref{lindblad3} without the impurity dissipators $\cD[\hat{A}_{\bm{r}}](\cdot)$. It is clear that the novel steady state persists even with these relaxation processes.

\section{Further figures}

\begin{figure*}
\centering
\subfloat[\label{subfig:tbsys-lrat}]{
    \includegraphics[scale=1.3,trim={0 2em 0 2em}, clip]{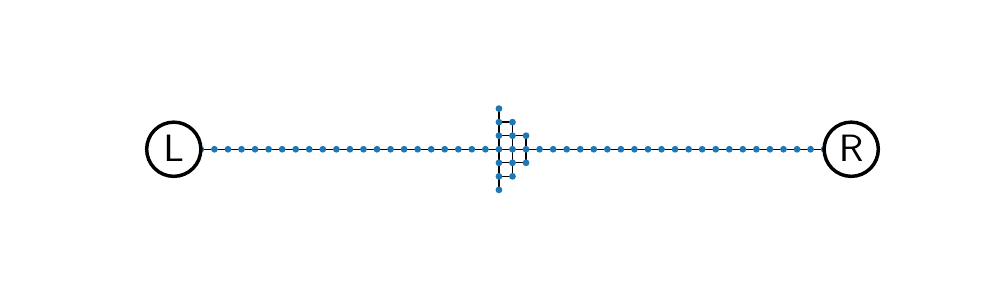}}\\
\subfloat[\label{subfig:tbsys-trat}]{
    \includegraphics[scale=1.3,trim={0 2em 0 2em}, clip]{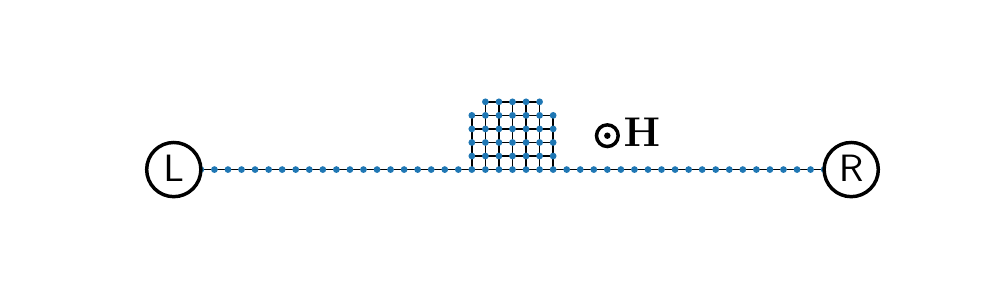}}\\
\subfloat[\label{subfig:tbsys-amzi}]{
    \includegraphics[scale=1.3,trim={0 2em 0 2em}, clip]{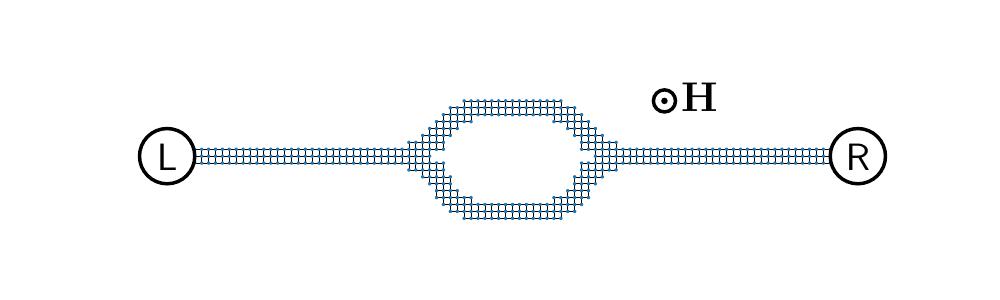}}\\
\subfloat[\label{subfig:tbsys-sdot}]{
    \includegraphics[scale=1.3,trim={0 2em 0 2em}, clip]{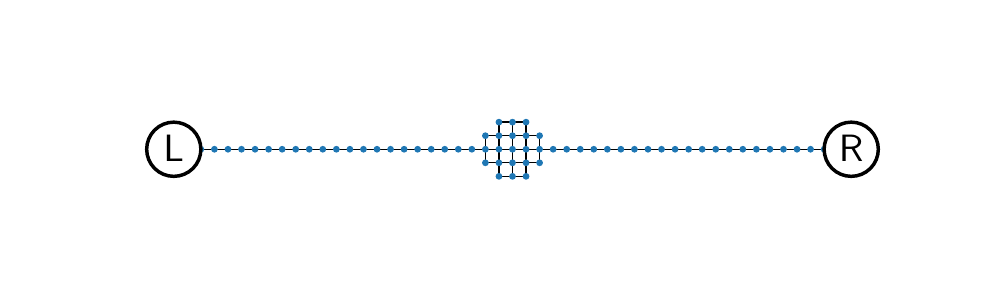}}
\caption{Layout of the devices sketched in \figref{dwell} as modeled numerically. The dots present the sites used in the tight-binding model.}
\label{fig:tbsys}
\end{figure*}

\begin{figure*}
\centering
\subfloat[\label{subfig:sdot-average-map}]{
    \includegraphics[scale=1]{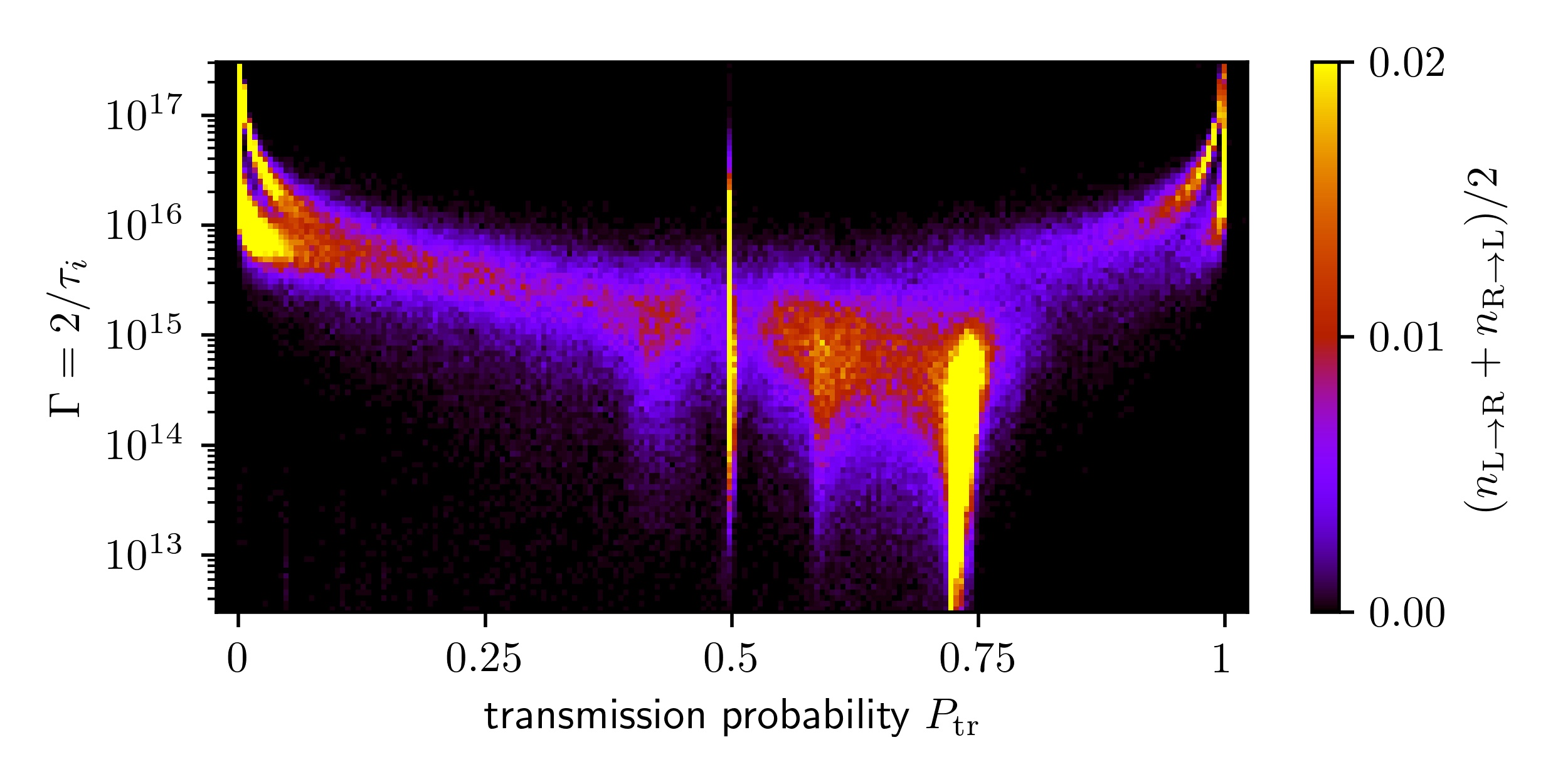}}\\
\subfloat[\label{subfig:sdot-diff-map}]{
    \includegraphics[scale=1]{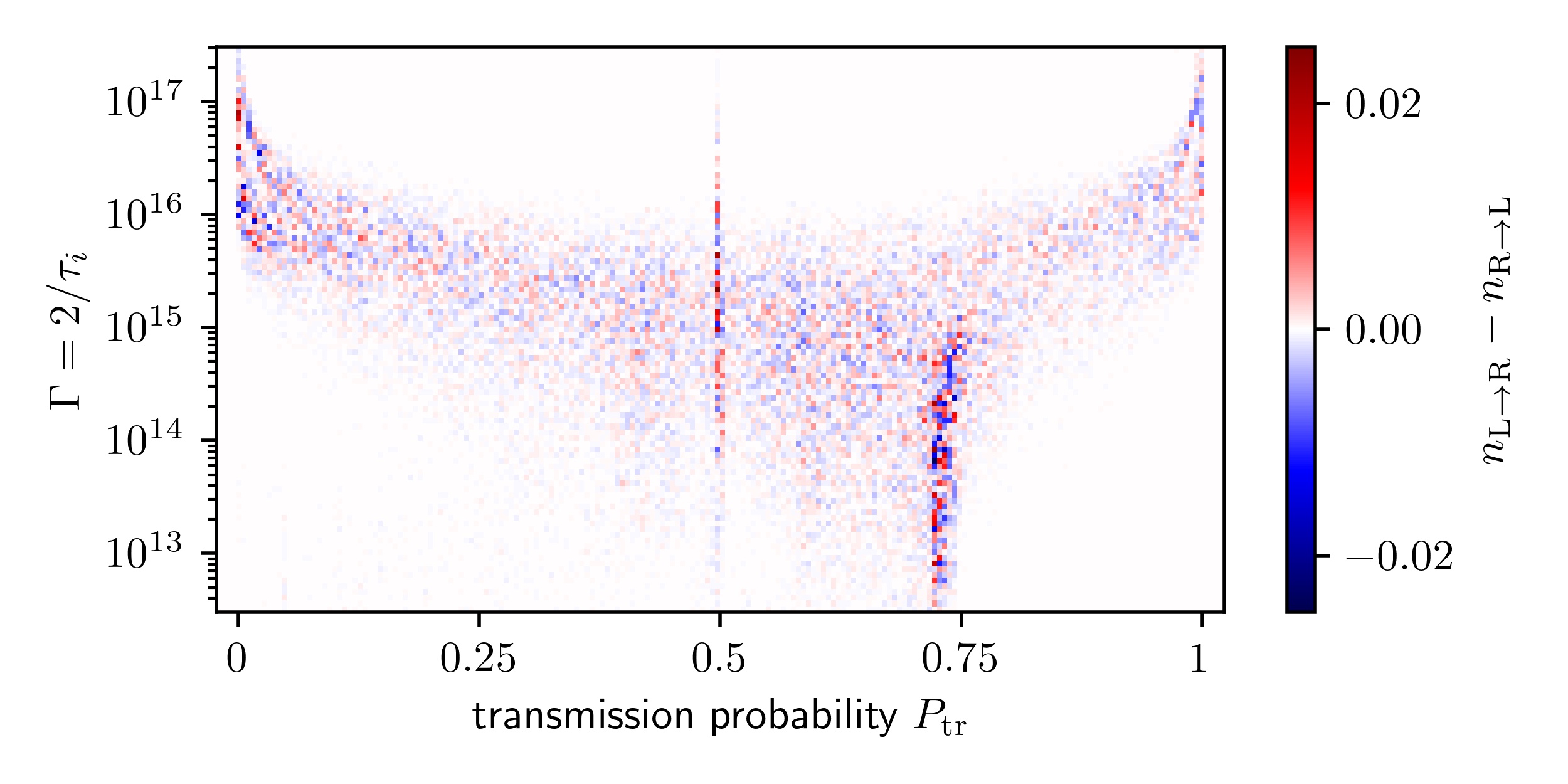}}
\caption{Transmission probability histograms calculated as a function of the decoherence rate $\Gamma$ for the symmetric device as shown in Figs. \ref{subfig:sys-sdot}, \ref{subfig:tbsys-sdot} and the inset. Nine localization centers in the center of the device mediate the coupling to the environment. The central device has a length of \SI{2.4}{\nm}. The data of the two colormaps have been obtained from \sdotValidTrajectories{} trajectories. \subref{subfig:sdot-average-map} The average over both directions shows one single probability for small $\Gamma$, a broad range of probabilities in the transition regime and only
$\Ptr=0$ or $\Ptr=1$ in the classical limit of large $\Gamma$, comparable to Fig.~\ref{subfig:avg-colormap}. \subref{subfig:sdot-diff-map} However, the difference plot shows the statistical noise only, lacking any direction dependent features.}
\label{fig:colormaps-sdot}
\end{figure*}

\begin{figure*}
  \centering
  \includegraphics[scale=1]{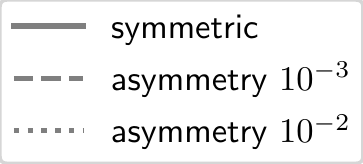}
  \subfloat[\label{subfig:lindblad-thermo-sdot-open}]{
    \includegraphics[scale=1]{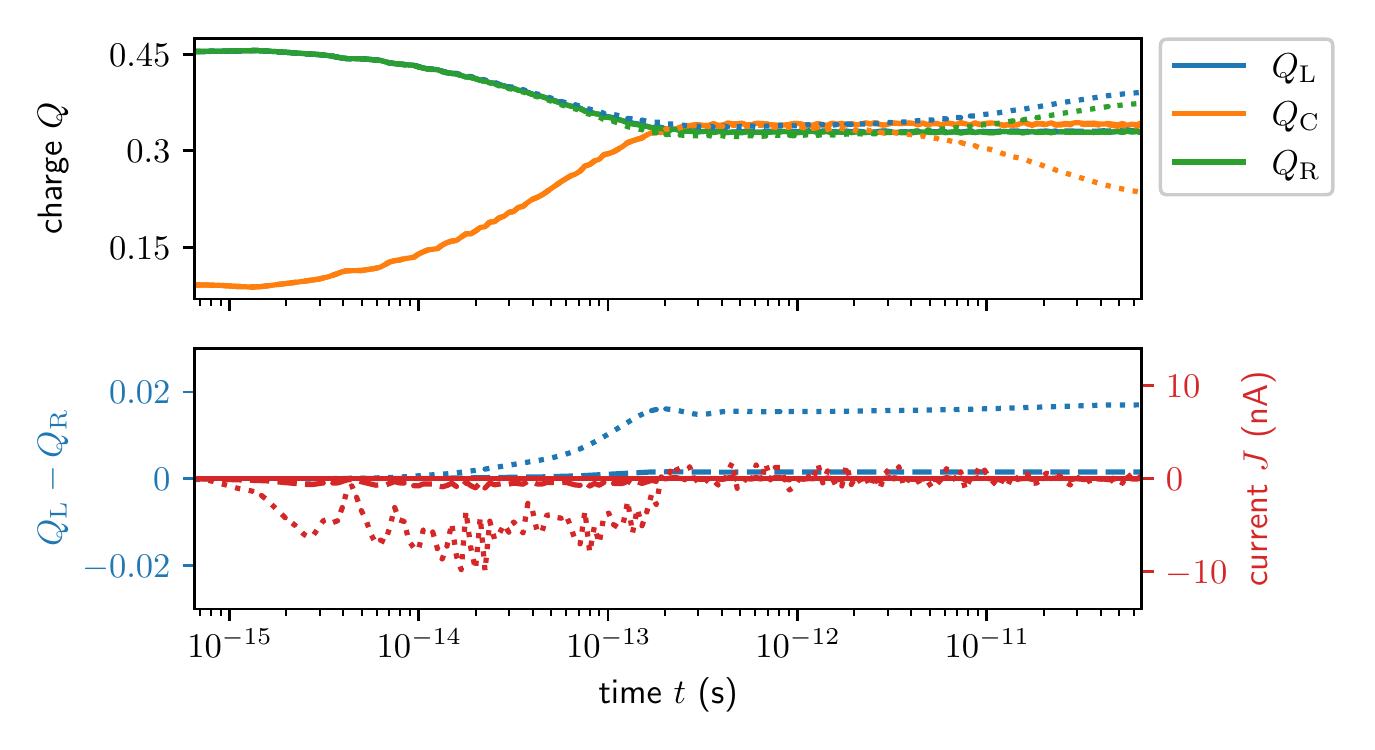}
  }\\
  \subfloat[\label{subfig:lindblad-thermo-sdot-closed}]{
    \includegraphics[scale=1]{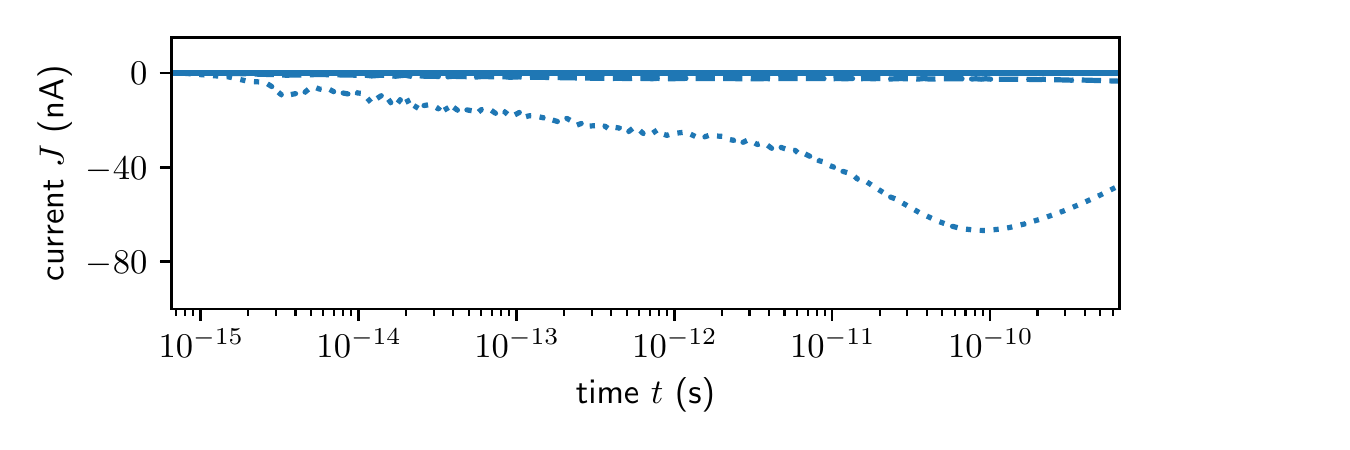}
  }
  \caption{Plots of the time evolution of charges in the open system as shown in \figref{lindblad-chain-sys}, but with fully symmetric (solid lines) and slightly asymmetric (dashed and dotted lines) potential barriers. The dashed line corresponds to a barrier similar to \eqnref{Vasym} but with the potentials $1.999t$, $2t$, $2.001t$ at sites \ket{4}, \ket{5}, \ket{6}. The dotted line corresponds to potentials $1.99t$, $2t$, $2.01t$ at sites \ket{4}, \ket{5}, \ket{6}.
  \subref{subfig:lindblad-thermo-sdot-open}~While no charge separation occurs in the fully symmetric open system, charge separation starts to occur with increasing barrier asymmetry.
  \subref{subfig:lindblad-thermo-sdot-closed}~No persistent current is flowing in the fully symmetric shorted system. Again, with increasing barrier asymmetry, persistent currents occur.
  The calculational method used is completely identical with the one used in \figref{lindblad-thermo}, the only difference being the symmetry of the barrier. Some of the blue ${Q_\mathrm{L}}$ curves are not visible because they fully overlap with the corresponding green ${Q_\mathrm{R}}$ curves.}
  \label{fig:lindblad-thermo-sdot}
\end{figure*}

Figures~\ref{fig:tbsys},~\ref{fig:colormaps-sdot} and \ref{fig:lindblad-thermo-sdot} further illustrate the time evolution of the electron waves in several devices.

\begin{figure*}
  \centering
  \includegraphics[scale=1]{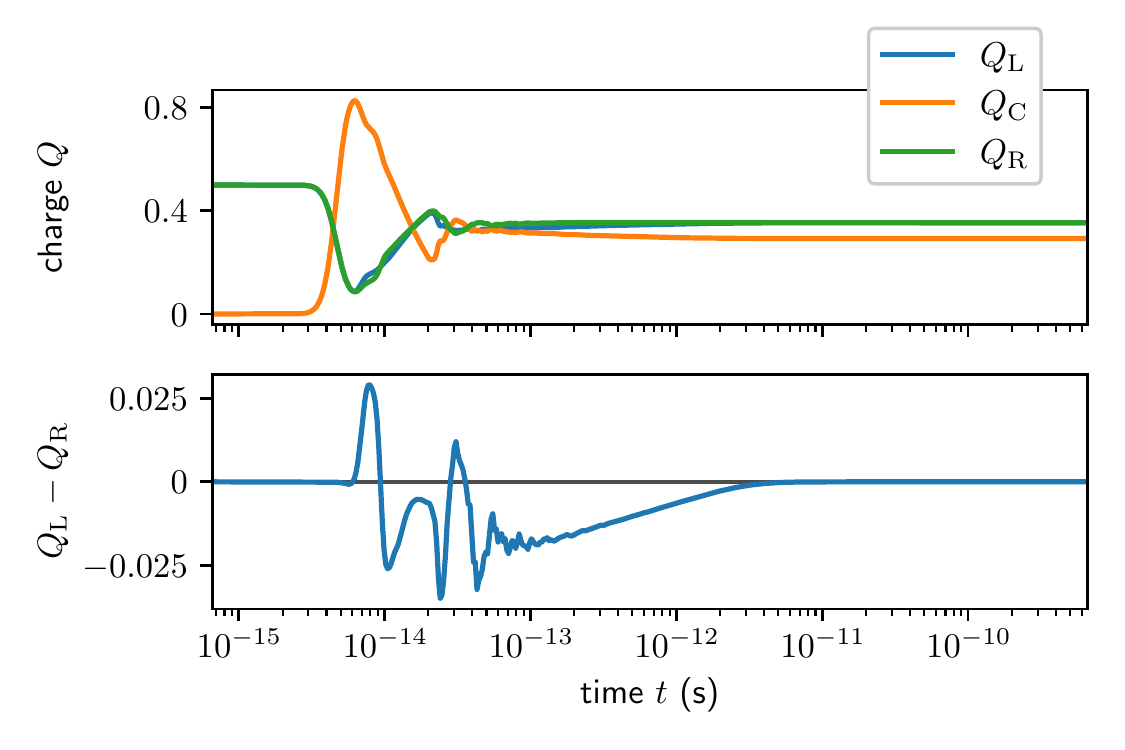}
  \caption{Plots showing the time evolution of charges in the device with longitudinal asymmetry (\subfigref{sys-lrat}) as calculated using \eqnref{lindblad2}. The upper panel shows the total charge $Q_\mathrm{L}$ in the left lead, in the right lead $Q_\mathrm{R}$, and within the central device $Q_\mathrm{C}$. The bottom panel shows the charge difference between the left and right leads, which is zero for large and small times. For intermediate times, however, the difference is finite, revealing a transient charge separation.}
  \label{fig:lindblad-steady-pack}
\end{figure*}

Figure~\ref{fig:lindblad-thermo-sdot} illustrates that the time evolution of the open and closed chains (Fig.~\ref{fig:lindblad-chain-sys}) for various degrees of asymmetry. As shown by the figure, the device asymmetry is mandatory for charge separation or persistent currents to occur. Figure~\ref{fig:lindblad-steady-pack} illustrates for the case of a device with longitudinal asymmetry that a a hermitean jump-operator does not generate a charge-separated steady-state.

\begin{figure*}
  \centering
  \includegraphics[scale=1]{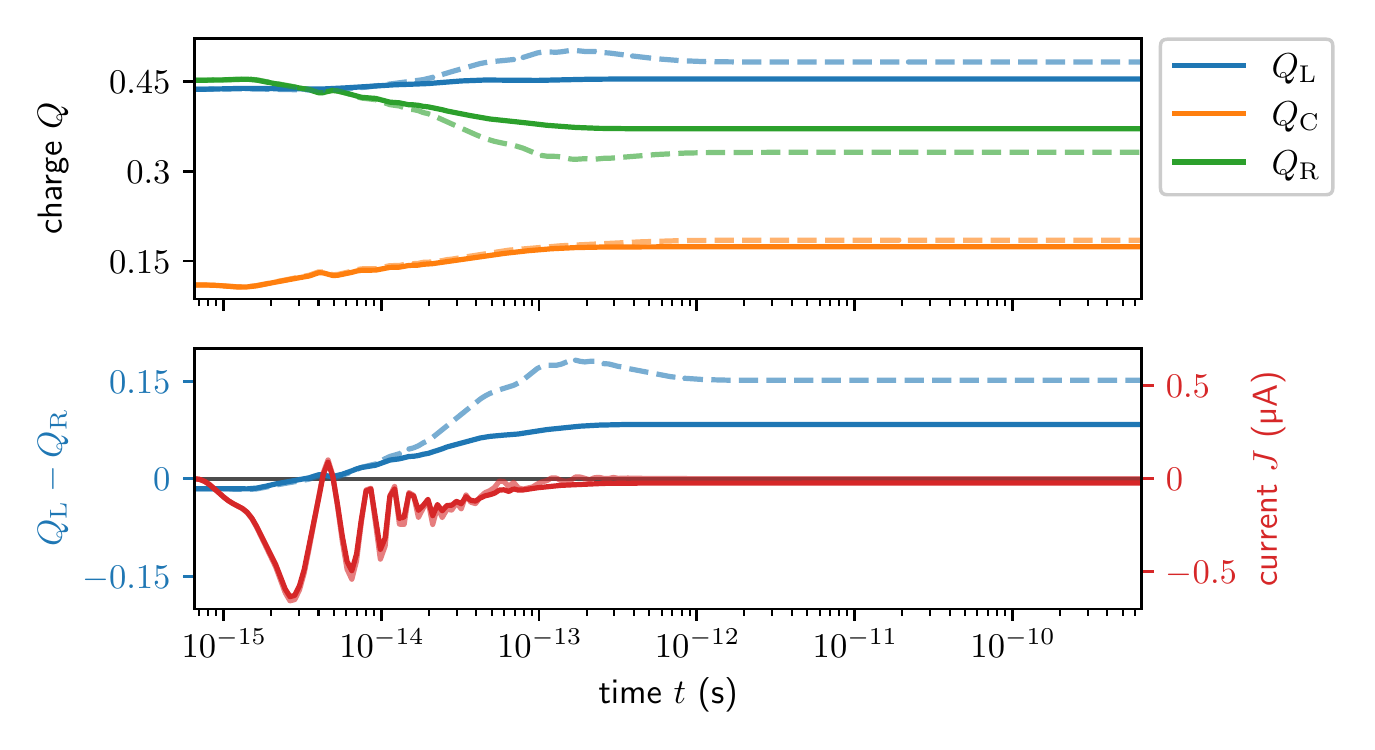}\\[-5.5cm]
  \hspace{12cm}\includegraphics[scale=1]{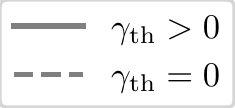}\\[4cm]
  \caption{Plots showing the time evolution of charges and currents in the open 9-site chain with additional jump operators that are specifically chosen to drive the system towards the state given by the thermal density matrix (see \ref{sec:minisys}). The rate ${\gamma}$ of the wave-packet generating jump operators and the rate ${\gamma_\mathrm{th}}$ of the additional thermalizing jump operators are given by ${\gamma=\SI{1.5e15}{Hz}}$ and ${\gamma_\mathrm{th}=\SI{3e14}{\Hz}}$, respectively. The steady-state charge imbalance indicates that the novel steady state persists even in the presence of thermalizing processes. The dashed lines show the case without the additional jump-operators, which is the same data as shown in \subfigref{lindblad-init-therm}.}
  \label{fig:lindblad-thermal-jumps}
\end{figure*}

Figure~\ref{fig:lindblad-thermal-jumps} shows that a charge-separated steady state is generated by the open 9-site chain even if additional jump operators that drive the system to thermal equilibrium are present in the Lindblad master equation.

\end{document}